\documentclass[fleqn, usenatbib]{mnras}

\usepackage{newtxtext}

\usepackage[T1]{fontenc}
\usepackage{ae,aecompl}

\usepackage{graphicx}	% Including figure files
\usepackage{amsmath}	% Advanced maths commands
\usepackage{amssymb}	% Extra maths symbols
\usepackage{subfig}
\usepackage[font=footnotesize]{caption}
\usepackage{wrapfig}
\usepackage{gensymb}
\usepackage{multirow}
\usepackage{tabularx}
\usepackage{csquotes}
\usepackage{float}
\usepackage{stackengine}
\usepackage[dvipsnames]{xcolor}
\usepackage{tablefootnote}
\usepackage{enumitem}
\usepackage{longtable}
\usepackage{threeparttablex}
\usepackage{hhline}
\usepackage{bm}		% Bold maths symbols, including upright Greek
\usepackage{pdflscape}	% Landscape pages
\usepackage{comment}

\def\maxififth{MAXI~J1535--571}
\def\maxithirt{MAXI~J1348--630}
\def\maxieight{MAXI~J1820$+$070}
\def\xte{XTE~J1550--564}
\def\hh{H1743--322}

\def\grs{GRS~1915$+$105}

\title[Decelerating jets from \maxithirt{}]{Modeling the kinematics of the decelerating jets from the black hole X-ray binary MAXI J1348--630}

\author[F. Carotenuto et al.]{F. Carotenuto,$^{1}$\thanks{E-mail: francesco.carotenuto@cea.fr}
A. J. Tetarenko,$^{2,3 \thanks{NASA Einstein Fellow}}$
S. Corbel$^{1,4}$
\\
$^{1}$AIM, CEA, CNRS, Universit\'{e} de Paris, Universit\'{e} Paris-Saclay, F-91191 Gif-sur-Yvette, France\\
$^{2}$East Asian Observatory, 660 N. A'\!oh$\bar{o}$k$\bar{u}$ Place, University Park, Hilo, Hawaii 96720, USA\\
$^{3}$Department of Physics \& Astronomy, Texas Tech University, Lubbock, TX 79409-1051, USA\\
$^{4}$Station de Radioastronomie de Nan\c cay, Observatoire de Paris, PSL Research University, CNRS, Univ. Orl\'eans, 18330 Nan\c cay, France\\
}
\date{Accepted XXX. Received YYY; in original form ZZZ}

\pubyear{2022}

\begin{document}
\label{firstpage}
\pagerange{\pageref{firstpage}--\pageref{lastpage}}
\maketitle

% Abstract of the paper
\begin{abstract}
\noindent
Black hole low mass X-ray binaries (BH LMXBs) can launch powerful outflows in the form of discrete ejecta. Observing the entire trajectory of these ejecta allows us to model their motion with great accuracy, and this is essential for measuring their physical properties. In particular, observing the final deceleration phase, often poorly sampled, is fundamental to obtain a reliable estimate of the jet's energy. During its 2019/2020 outburst, the BH LMXB \maxithirt{} launched a single-sided radio-emitting jet that was detected at large scales after a strong deceleration due to the interaction with the interstellar medium (ISM). We successfully modelled the jet motion with a dynamical external shock model, which allowed us to constrain the jet initial Lorentz factor $\Gamma_0 = 1.85^{+0.15}_{-0.12}$, inclination angle $\theta = 29.3\degree_{-3.2\degree}^{+2.7\degree}$ and ejection date $t_{\rm ej} = 21.5_{-3.0}^{+1.8}$ (MJD -- $58500$). Under simple assumptions on the jet opening angle and on the external ISM density, we find that the jet has a large initial kinetic energy $E_0 = 4.6^{+20.0}_{-3.4} \times 10^{46}$ erg, far greater than what commonly measured for LMXBs from the jet's synchrotron emission. This implies that discrete ejecta radiate away only a small fraction of their total energy, which is instead transferred to the environment. The jet power estimate is larger than the simultaneous available accretion power, and we present several options to mitigate this discrepancy. We infer that \maxithirt{} is likely embedded in an ISM cavity with internal density $n = 0.0010^{+0.0005}_{-0.0003}$ cm$^{-3}$ and radius $R_{\rm c} = 0.61^{+0.11}_{-0.09}$ pc, which could have been produced by the system's previous activity, as proposed for other BH LMXBs.

\end{abstract}

% Select between one and six entries from the list of approved keywords.
% Don't make up new ones.

%%%%%%%%%%%%%%%%%%%%%%%%%%%%%%%%%%%%%%%%%%%%%%%%%%

%%%%%%%%%%%%%%%%% BODY OF PAPER %%%%%%%%%%%%%%%%%%

\section{Introduction}
\label{sec:Introduction}

Accreting black holes (BH) in low mass X-ray binaries (XRBs) spend most of the time in a steady, quiescent state and enter into sporadic outburst phases during which their accretion rate and luminosity dramatically increase over timescales of weeks to months \citep{Tetarenko_2016, Corral_santana}. While in outburst, BH XRBs cycle through a number of accretion states that are characterized by very different X-ray spectral and timing properties, allowing us to study the same object over the full range of accretion regimes on human timescales (e.g. \citealt{Fender_belloni_gallo, Remillard_xrb,  Belloni_Motta_2016}). In these systems, also called \textit{microquasars}, a significant fraction of the disk mass, not advected into the BH event horizon, is channeled into powerful, relativistic jets. Self-absorbed synchrotron radiation from collimated, compact jets is generally detected in the hard X-ray state, as these outflows are observed from the radio to near-infrared bands \citep{Corbel_2000, Fender_2001}, and possibly up to the optical band \citep{Russell_2006}. Such jets are strongly coupled to the inner accretion flow and respond to changes in the accretion rate of the system, as deduced from the strong correlation that exists between the radio and X-ray luminosities of BH XRBs in the hard state (e.g. \citealt{Hannikainen_1998, Corbel_2000, Gallo_2003, Carotenuto_lrlx}).

Perhaps the most spectacular feature of BH LMXBs is the launch of discrete ejecta, which are bipolar blobs of plasma moving away from the core, often displaying apparent superluminal motion, similarly to what is observed in AGN (e.g.\ \citealt{Marscher_2002, Gomez_2008}). While all BH LMXBs are thought to produce jets, resolved discrete ejecta have been observed in radio in a dozen of sources so far (e.g. \citealt{Mirabel1994, GRO1655, Fender1999, Mioduszewski, Gallo_2004, Yang2010, Rushton, Russell_1535, Miller-Jones2019, Bright, Carotenuto2021}). For some of them both the approaching and receding components have been detected, while for a large fraction of sources only one component (usually the approaching one, due to Doppler boosting) was observed. Discrete ejecta have been observed also in neutron star LMXBs (e.g. Sco X-1 and Cir X-1, \citealt{Fomalont_2001, Calvelo_2012}).

After launch, the ejecta display a first phase of adiabatic expansion, producing strong radio flares and appearing to leave the system at constant, high velocities (close to $c$). Throughout their evolution, the plasma bubbles emit synchrotron radiation at radio wavelengths and their optical depth decreases as the physical size increases \citep{vanderlaan, Fender_2019_equipartition}, up to displaying an optically thin spectrum, with a spectral index $\alpha \simeq -0.6$ (e.g.\ \citealt{Corbel2002_xte}), where the radio flux density follows $S_{\nu} \propto \nu^{\alpha}$. 

However, for the majority of the discrete ejecta observed so far, the jet motion has only been partially observed, mostly focusing on the initial motion after ejection, in which we usually resolve at small scales (i.e.\ close to the BH) the ejecta travelling at constant speeds.
But after this first phase, the jets can travel unseen for months to years, before being detected again at large (parsec) scales, with the emission coming from their interaction with the surrounding ISM. Such interaction causes the jets to strongly decelerate and produces broadband (from radio to X-rays) synchrotron radiation from \textit{in-situ} particle acceleration, up to TeV energies (e.g.\ \citealt{Corbel2002_xte, Espinasse_xray}).
Decelerating discrete ejecta at large scales have been so far only detected in a handful of sources \citep{Tomsick_2003, Kaaret_2003, Corbel2005_h17, Miller-Jones_2007, Yang2010, Miller_jones_sedov, Russell_1535, Espinasse_xray, Carotenuto2021}.
However, following the jet deceleration phase is turning out to be particularly important to improve our understanding of their production, energetics, dynamics and impact on the surrounding environment. Covering the last phase of the jets evolution allows us to obtain a complete picture of their motion, and this in turn enables us to fully model their evolution with physical models (e.g.\ \citealt{Steiner_xte}). It is then possible to infer the jet physical parameters, such as the true jet speed, energy, inclination angle and mass. In particular, the jet initial energy is a key parameter, as today the total energy budget of these jets is unknown. Constraints on the minimum energy can be obtained by considering the synchrotron emission at the initial radio flares \citep{Longair, Fender_2019_equipartition}. However, such estimates (often between $10^{39}$ and $10^{42}$ erg) are likely to largely underestimate the total energy content of the jet \citep{Steiner_xte, Bright}. In addition, there is large uncertainty on the jet matter content (e.g.\ \citealt{Romero_2017}), and the knowledge of the jet mass and energy can be particularly useful for understanding the jet composition.

Modeling the jet motion is also of prime importance to precisely constrain the time of ejection, which is fundamental to put the transient jet launch in context with the other multi-wavelength observational signatures and to have a comprehensive view of the source evolution during the state transition. The ejection time can already be roughly inferred from the radio flaring behaviour, which is the prominent signature of an ejection, but the unknown delay between the effective ejection and the subsequent flare, opacity effects, and the eventually sparse observational sampling, can limit the accuracy on the estimation of the ejection time (e.g.\ \citealt{Russell_1535}). Radio flares are usually observed at the transition between the hard and soft intermediate states (HIMS to SIMS) on the top branch of the Hardness-Intensity diagram \citep{Corbel_2004, Fender_belloni_gallo}, 
but it is still not clear what triggers (and powers) the ejection and if there is a unique time-ordered combination of observational signatures associated with discrete ejections. 
The onset of the transient jet might be a direct consequence of the compact jet quenching in the HIMS \citep{Russell_quenching}, but other explanations are possible (e.g.\ \citealt{Kaiser, Vadawale_2003, Rodriguez_2003}).
A significant change in the X-ray emission properties is generally observed during the IMS, consisting of a sharp decrease in the fractional rms variability (e.g. \citealt{Belloni_2005}) and the sudden appearance of Type-B Quasi Periodic Oscillations (QPOs, \citealt{Soleri2008}). While strongly suggested in some cases (e.g.\ \citealt{Homan_qpo}), such connection has never been clearly confirmed, and the observations of several sources seem to suggest that this connection might not be universal (e.g. \citealt{Miller-Jones_h1743, Russell_1535, Wood_2021, Zhang_2021}).

The jet deceleration also encodes information on the environment surrounding the black hole, including the presence of under-dense ISM cavities possibly produced by previous activity of the system.
Such cavities were inferred to exist for the majority of BH LMXBs \citep{Heinz_2002, Hao} and their borders are assumed to be sites of significant jet/ISM interaction. Under several assumptions, modeling the jet motion can allow us to obtain hints on the local density and on the cavity geometry, as done for XTE J1550--564 \citep{Steiner_xte} and \hh{} \citep{Hao, Steiner_h17}.
Since it is not known \textit{a priori} when the jet will start its deceleration phase, a complete coverage of its motion requires dense and uniform radio monitoring campaigns, with a constant ($\sim$weekly) cadence and on timescales of months to years (e.g.\ \citealt{Migliori2017, Bright}).

As the jets decelerate in the ISM, they transfer large amounts of energy and matter into their surrounding environment, and thus they represent an important source of feedback. In addition to possibly carving out under-dense cavities and thus shaping the ISM distribution, it is speculated that jets from galactic XRBs could inject roughly 1\% average supernovae luminosity into the environment \citep{Fender_2005}, heating and shocking the surrounding ISM. Their feedback might so important that they could effectively stimulate star formation \citep{Mirabel_starform}. The jet feedback on the ISM can be studied and characterized at higher frequencies, as the altered local chemistry and the distribution of excited gas can be probed through the observation and mapping of molecular line emission, as for instance recently done with ALMA in the sub-mm band for \grs{} \citep{Tetarenko_alma}, GRS\ 1758--258 and 1E~1740.7--2942 (\citealt{Tetarenko_2020_ism}, although these two sources display persistent jets). Such interaction regions can be used to study in detail the jet properties (such as the total energy, e.g.\ \citealt{Gallo_2005}), but their identification is rather difficult at the present day. Therefore, tracking the ejecta motion and looking for evidence of deceleration can be a way to identify systems that are possibly embedded in cavities, which are promising jet-ISM interaction sites.

\subsection{\maxithirt{}}
\label{sec:maxithirt}

\maxithirt{} is a new BH LMXB discovered by the MAXI monitor on board the ISS \citep{Matsuoka_maxi} in January 2019 \citep{Yatabe2019, Tominaga_1348}, when it quickly became one of the brightest X-ray transients of 2019. During its 2019/2020 outburst, the source first completed a whole cycle in the Hardness Intensity Diagram (HID), with strong radio flaring at the first state transition \citep{Carotenuto_atel}, and then displayed a sequence of hard state re-brightenings that lasted until September 2020 (e.g. \citealt{DRussell_atel_opt_2, Yazeedi_atel_2, Pirbhoy_atel, Shimomukai_atel, Baglio_last_1348_atel}).

The orbital period of the system is currently unknown, and we do not have information on the companion star and on the BH mass or spin.
Observations of {\rm HI} absorption in \maxithirt{} carried out with the Australian Square Kilometre Array Pathfinder (ASKAP) and with MeerKAT yielded a distance of $D = 2.2^{+0.6}_{-0.5}$ kpc \citep{Chauhan2021}, making \maxithirt{} a relatively close system. A mild tension exists between this measurement and a distance of $\sim$3.3 kpc obtained with the eROSITA X-ray detection of a dust-scattering halo \citep{Lamer_2021}, which is discussed in Section \ref{sec:Dust-scattering halo distance}. 

In \cite{Carotenuto2021} we have presented the full X-ray and radio monitoring of the source during its 2019/2020 outburst, including data from the MeerKAT radio-interferometer \citep{Jonas2016, Camilo2018}, the Australia Telescope Compact Array (ATCA, \citealt{Frater_1992}), the XRT telescope \citep{Burrows_xrt} on-board the Neil Gehrels \emph{Swift} Observatory \citep{Gehrels} and the Monitor of All-sky  X-ray Image (MAXI, \citealt{Matsuoka_maxi}). Our radio observations covered the entire evolution of compact jets (including their rise, quenching and re-activation) through the different phases of the outburst. In a subsequent work on the hard state disk/jet connection, we found that \maxithirt{} belongs to the growing group of \textit{radio-quiet} or \textit{outlier} BH LMXBs on the radio/X-ray diagram \citep{Carotenuto_lrlx}.

\maxithirt{} has also been monitored with the Neutron Star Interior Composition Explorer (NICER) on board the ISS and the spectral-timing analysis of the fast X-ray variability has been recently conducted, with the observations of state transitions, the detection of different types of QPOs and the identification of a dominant hard Comptonised emission in the Type-B QPOs at the hard-to-soft state transition \citep{Zhang2020, Belloni_2021, Garcia_2021, Zhang_2021}.

\maxithirt{} displayed spectacular discrete ejecta \citep{Carotenuto2021}. Two single-sided jets were launched $\sim$2 months apart and displayed the highest proper motion ever measured for an accreting BH ($\gtrsim 100$ mas day$^{-1}$). After travelling with constant speed, the first jet component was detected at large scales ($\sim$26 arcsec from the core, corresponding to a distance of at least 0.3 pc) several months after the ejection, when it underwent a strong deceleration, covered in radio in great detail \citep{Carotenuto2021}.
The motion of these ejecta has already been presented and studied with the application of phenomenological models, which allowed us to qualitatively describe the jet evolution and to obtain a reliable estimate of the ejection date for both components.
In this paper we focus on the first discrete ejecta (labelled RK1 in \citealt{Carotenuto2021}) and we go one step further in the study of the motion of this jet by applying to our data a full dynamical model based on external shocks \citep{Wang_model}, in order to infer physical parameters of the jet and of the environment of \maxithirt{}. In Section \ref{sec:Data} we present the data considered in this paper, while in Section \ref{sec:The external shock model} we discuss the external shock model and 
in Section \ref{sec:Results} we present the results of the application of such model to our data. We then discuss our findings in relation to the current understanding of jets from XRBs in Section \ref{sec:Discussion} and we summarize our conclusions in Section \ref{sec:Conclusions}.

\section{Data}
\label{sec:Data}

We monitored the entire outburst of \maxithirt{} in radio with MeeKAT at 1.28 GHz, as part of the ThunderKAT Large Survey Programme \citep{ThunderKAT}, and with ATCA at 5.5 and 9 GHz. In X-rays, \maxithirt{} was regularly monitored by \textit{Swift}/XRT and by MAXI. The full observing campaign on \maxithirt{} has already been presented in \cite{Carotenuto2021} and we refer to that work for details on the radio and X-ray data processing. 

In this paper we focus on the observations of the first discrete ejecta from \maxithirt{}, that have been presented in \cite{Carotenuto2021}, where the jet is labelled RK1. We use the MeerKAT and ATCA detections of the first component to fit the entire jet motion, therefore we consider the position of the jet in every epoch in which it was detected and its angular separation with respect to the core position of \maxithirt{}, which is at R.A $= 13^{\rm h} 48^{\rm m}12.79^{\rm s} \pm 0.01^{\rm s}$ and Dec $= -63\degree16\arcmin28.6\arcsec \pm 0.2\arcsec$ (J2000, \citealt{Carotenuto2021}).

The coordinates of the jet at each epoch are reported in Table \ref{tab:first_jet_angsep}, for which data from MeerKAT and from ATCA at 5.5 GHz were used. The jet angular separation is computed as the great circle distance between the ejecta and the \maxithirt{} position, either the fitted one in case of core detection on the same epoch, or the reference core position mentioned above in case of core non-detection. In the first case, we are not affected by global systematics in the position error estimation. After a first part of ballistic, high-speed motion that started roughly between MJD 58518 and 58523, the deceleration of RK1 around MJD 58775 was rather abrupt, a phenomenon not observed in the majority of discrete ejecta from BH LMXBs (e.g\ \citealt{Mirabel1994, Miller-Jones_h1743, Rushton, Miller-Jones2019}). 

\section{The external shock model}
\label{sec:The external shock model}

The jet dynamical model that we adopt in this work was originally developed by \cite{Wang_model}, and later expanded and applied to the discrete ejecta of \xte{} and \hh{} \citep{Hao, Steiner_xte, Steiner_h17}. While in principle the model was designed for gamma-ray bursts afterglows, it can describe the evolution of mildly relativistic jets from XRBs, and we present it again for clarity. In particular, we consider the most recent implementation of \cite{Steiner_xte}.

We consider a pair of symmetric ejecta launched at the same time in opposite directions at an inclination angle $\theta$ with respect to the line of sight, with an initial Lorentz factor $\Gamma_0$ and kinetic energy $E_0$. The jets take the shape of confined, conical beams and expand with a constant half-opening angle $\phi$ inside an ambient medium with constant density $n$.
As the jets expand, they sweep up surrounding material and transfer their kinetic energy to the internal energy of the entrained material through external shocks, heating the ISM and consequently undergoing deceleration. Radiation losses 
at the shock front are assumed to be negligible and hence the jet expansion can be considered adiabatic throughout the whole evolution. Considering one of the two ejected components, it is therefore possible to write the energy conservation equation, following \cite{Huang_1999} and \cite{Wang_model}
\begin{equation}
E_0 = (\Gamma -1) M_0 c^2 + \sigma(\Gamma_{\rm sh}^2-1)m_{\rm sw}c^2
\label{eq:energy}
\end{equation}
where the first term on the right-hand side is the kinetic energy of the ejecta and the second term is the internal energy of the shock. Here $\Gamma$ is instantaneous jet bulk Lorentz factor, $M_0$ is the jet mass and $\Gamma_{\rm sh}$ is the Lorentz factor at the shock front, while $\sigma$ is a numerical factor equal to $6/17$ for ultra-relativistic shocks and $\sim$0.73 for non-relativistic shocks \citep{Blandford_mckee_1976}. A numerical scaling can be adopted to interpolate between the two regimes \citep{Huang_1999, Wang_model, Steiner_xte}
\begin{equation}
\sigma=0.73-0.38\beta
\label{eq:sigma}
\end{equation}
where $\beta=(1-\Gamma^{-2})^{1/2}$. Moreover, $m_{\rm sw}$ is the mass of the swept-up material, and it can be written as
\begin{equation}
m_{ \rm sw} = \frac{\phi^2 \pi m_{\rm p} n R}{3}
\label{eq:mass_sweptt}
\end{equation}
where $R$ is the distance from the core. A forward shock develops at the contact discontinuity between the jet and the ISM. Such shock continuously heats the encountered ISM and randomly accelerates particles. By using the jump conditions for an arbitrary shock, it is possible to express the shock Lorentz factor as a function of the jet bulk Lorentz factor \citep{Blandford_mckee_1976, Steiner_xte}
\begin{equation}
\Gamma_{\rm sh}^2=\frac{(\Gamma+1)(\hat{\gamma}(\Gamma-1)+1)^2}{\hat{\gamma}(2-\hat{\gamma})(\Gamma-1)+2}
\label{eq:Lorentz_shock}
\end{equation}
where $\hat{\gamma}$ is the adiabatic index that varies between $4/3$ (ultra-relativistic shocks) and $5/3$ (non-relativistic shocks). Following \cite{Steiner_xte}, we interpolate between the two limits with
\begin{equation}
\hat{\gamma}=\frac{4\Gamma+1}{3\Gamma}.
\label{eq:interpolation}
\end{equation}

Due to the possible evidence of BH LMXBs being harbored in low-density ISM cavities \citep{Hao}, including \maxithirt{} (see Section \ref{sec:Motion in a low-density ISM cavity} and \citealt{Carotenuto2021}), we include such cavity in the model, modifying the expression for the swept-up mass. 
Following the example of \cite{Steiner_xte}, we let the jet first expand in an under-dense environment parametrized with a radius $R_{\rm c}$ and a density jump $\delta$ at the cavity wall, so that, for a standard ISM density $n_{\rm ism}$ outside, we have a density  $n= n_{\rm ism}/\delta$ inside. The entrained mass then becomes
\begin{equation}
  m_{\rm sw}= \frac{\phi^2 \pi m_{\rm p} n_{\rm ism}}{3 \delta} \times \begin{cases}
              R^3, &R\leq R_{\rm c},\\
              R_{\rm c}^3 +\delta[R^3-R_{\rm c}^3], &R> R_{\rm c},
            \end{cases}
\label{eq:final_mass}
\end{equation}
This scenario is consistent with a jet that travels first at constant and high speed in a low-density environment, which constitutes a large scale ISM cavity around the system, before hitting the border of the cavity, decelerating and producing radio emission from the subsequent external shocks between the jet material and the higher-density ISM encountered, as proposed by \cite{Hao}.

The relativistic kinematic equation for the approaching component is \citep{Rees_1966, Mirabel1994}
\begin{equation}
\frac{dR}{dt}=\frac{\beta c}{1 - \beta\cos{\theta}}
\label{eq:proper motion}
\end{equation}
and the measurable projected angular distance is
\begin{equation}
\alpha(t)=\frac{R(t) \sin{\theta}}{D}
\label{eq:alpha}
\end{equation}
where $D$ is the source distance.

It is possible to obtain the proper motion of the jet on the plane of the sky for every set of the nine parameters that compose the model: 
the jet initial kinetic energy $E_0$ and Lorentz factor $\Gamma_0$, the jet axis inclination angle $\theta$ and half-opening angle $\phi$, the source distance $D$, the cavity radius $R_{\rm c}$, the density jump $\delta$, the external ISM density $n_{\rm ism}$ and the ejection time $t_{\rm ej}$.
However, a degeneracy exists in this model between the three parameters $E_0, \phi$ and $n_{\rm ism}$, as they appear as a single term in Equation \ref{eq:energy}, so that only the combination $E_0/n_{\rm ism} \phi^2$ can be effectively constrained. In order to be able to have an estimate of the initial energy, we fix the external density to the canonical value $n_{\rm ism} = 1$ cm$^{-3}$ and we choose the reasonable value for the half-opening angle of $\phi = 1\degree$, \citep{Kaaret_2003, Miller-jones2006}, which is also consistent with the observational constraints from \cite{Carotenuto2021}.

To obtain a jet trajectory for every set of parameters, we integrate Equation \ref{eq:proper motion} starting at a time $t_{\rm ej}$ from a distance $R_0 = 10^8$ cm (the results depend very weakly on the choice of $R_0$), by numerically solving at every time step Equation \ref{eq:energy} for the instantaneous jet Lorentz factor, and then we convert the distance traveled by the jet to the angular separation $\alpha$ (Equation \ref{eq:alpha}), in order to compare the obtained angular separation curve with our data (see Table \ref{tab:first_jet_angsep}).

\begin{figure*}
\begin{center}
\includegraphics[width=\textwidth]{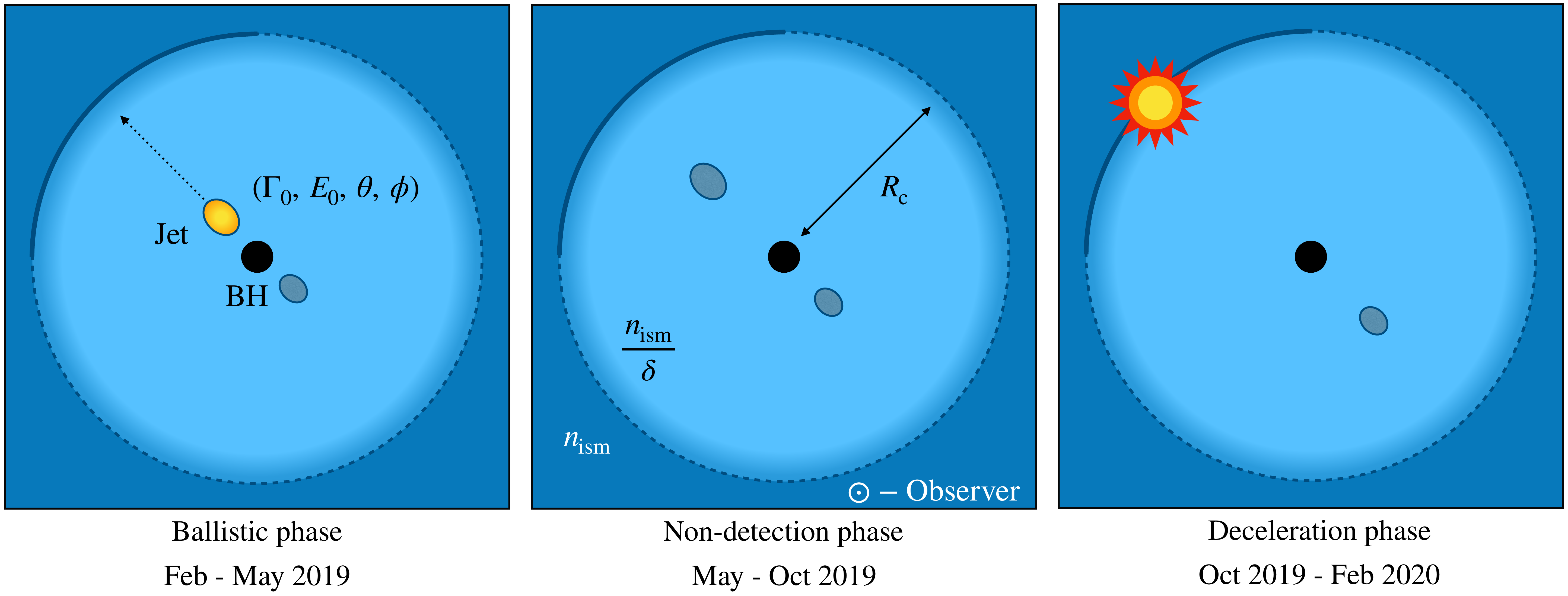}
\caption{Schematic representation of the observed jet behaviour and proposed scenario for \maxithirt{}, along with the parameters of the model adopted in this work. The line of sight is perpendicular to the plane of the drawing. The dashed line at the cavity border represents the fact that the cavity spherical geometry is an assumption of this scheme, and not of the model itself.
The jet is characterized by the four parameters $\Gamma_0, E_0, \theta, \phi$ (initial bulk Lorentz factor, kinetic energy, inclination angle and half-opening angle of the jet, respectively, see Section \ref{sec:The external shock model}) and displays a first phase of constant speed motion in an environment with density $n_{\rm ism}/\delta$. After a period of non-detection, the jet reaches the cavity wall at a distance $R_{\rm c}$ from the BH and transitions to a Sedov evolutionary phase. The unseen counter-jet moving in the south-west direction is also shown.}
\label{fig:cavity}
\end{center}
\end{figure*}

Since for \maxithirt{} only single-sided jets were detected, for the model application in this work we only use the angular distance information and not the flux ratios between the approaching and receding components, as for instance done for \xte{} and \hh{} (e.g.\ \citealt{Wang_model, Hao, Steiner_xte}), because the upper limits from the receding component do not allow us to obtain additional significant constraints on the physical parameters of the jet or of the cavity.
In particular, this is the first time in which the external shock model is applied to a single-sided BH XRB jet. A schematic representation of the proposed scenario is shown in Figure \ref{fig:cavity}.

\section{Results}
\label{sec:Results}

\setlength{\tabcolsep}{8pt}
\setlength{\extrarowheight}{.7em}
\begin{table*}
\caption{Parameters of the external shock model applied in this work, chosen priors with their distribution and allowed values, and fit results with and without the inclusion of a low-density ISM cavity. The values quoted are the median parameter and the $1\sigma$ confidence intervals derived from the MCMC run. We also report the BIC for the two cases.}
\label{tab:fit_params_jets}
\begin{tabular}{*{6}{c l l c c c}}
\hline
\hline
Parameter        & Description 							  & Prior distribution 							   & Prior interval & Uniform density        & ISM cavity\\
\hline
$\Gamma_0$       & Initial bulk Lorentz factor at launch  & Uniform 								       & $1$--$100$         & $1.85_{-0.16}^{+0.25}$  & $1.85_{-0.12}^{+0.15}$\\
$\log_{10}{E_0}$ & Initial kinetic energy at launch (erg) & Uniform 									   & $35$--$55$         & $48.65_{-0.47}^{+0.51}$ & $46.66_{-0.60}^{+0.73}$\\
$\theta$         & Inclination angle (\degree) 			  & Truncated normal ($\mu = 30$, $\sigma = 20$)   & $0$--$90$          & $28.31_{-3.40}^{+4.77}$ & $29.27_{-3.23}^{+2.71}$\\
$R_{\rm c}$      & Cavity radius (pc) 					  & Uniform 								       & $0$--$2$           & $0$ (fixed)               & $0.61_{-0.09}^{+0.11}$\\
$\delta$         & Density jump at cavity wall 			  & Uniform in $\log{\delta}$ 					   & $1$--$10^4$      & $1$ (fixed)               & $980_{-359}^{+514}$\\
$D$ 			 & Distance (kpc) 						  & Truncated normal ($\mu = 2.2$, $\sigma = 0.6$) & $1$--$8$           & $2.38_{-0.35}^{+0.52}$  & $2.46_{-0.27}^{+0.31}$\\
$t_{\rm ej}$     & Ejection date (MJD -- $58500$) 		  & Uniform 									   & $0$--$23.5$        & $22.92_{-2.51}^{+1.35}$ & $21.45_{-3.02}^{+1.83}$\\
\hline
BIC & Model selection & \ \ \ \ \ \ -- & -- & $120.4$ & $49.5$\\
\hline
\end{tabular}
\end{table*}

We fit our data with the external shock model presented in Section \ref{sec:The external shock model} with a Bayesian approach, applying a Monte Carlo Markov Chain (MCMC) code implemented with the 
\textsc{emcee} package \citep{emcee_paper}.
For every point of the parameter space, Equation \ref{eq:proper motion} was integrated using {\tt odeint} from the SciPy package \citep{2020SciPy-NMeth}.

We include the maximum amount of available information in the choice of our priors, which are physically motivated from our knowledge of \maxithirt{} and BH XRBs in general. Flat priors are assumed in case we do not have a preferred expectation value for a specific parameter. The prior distributions used for all of our parameters are listed in Table \ref{tab:fit_params_jets}. We adopted a flat prior for $\Gamma_0$ and a log-flat prior for $E_0$, with a large range of allowed values. For $\theta$, we used a normal distribution (truncated outside the interval 0\degree--90\degree) centered on 30\degree and with a sigma of 20\degree, taking into account the observational constraints from \cite{Carotenuto2021} and the fact that $\theta$ is unlikely to be small (i.e.\ the system is unlikely to be almost face-on). We then adopted a flat prior between 0 and 2 pc for $R_{\rm  c}$ and a log-flat prior for $\delta$, with boundaries at 1 and 10$^4$, including the whole range of inferred values for other BH LMXBs (e.g.\ \citealt{Heinz_2002, Hao, Steiner_xte}). A truncated normal is chosen for $D$, and it is centered on 2.2 kpc with a sigma of 0.6 kpc \citep{Chauhan2021}, while the prior on $t_{\rm ej}$ is flat and truncated at MJD 58523.5, since the jet cannot be ejected after having produced the strong radio flare observed with MeerKAT at that time ($\sim$0.5 Jy, \citealt{Carotenuto2021}).

Every MCMC run was conducted using 110 walkers. For each run, we consider that convergence is reached when the positions of the walkers in the parameter space are no longer significantly evolving. 
Once the chains have converged, the best fit result for each parameter is taken as the median of the one-dimensional posterior distribution obtained from the converged chains, while the 1$\sigma$ uncertainties are reported as the difference between the median and the 15th percentile of the posterior (lower error bar), and the difference between the 85th percentile and the median (upper error bar). The best fit results for our runs are shown in Table \ref{tab:fit_params_jets}.
For choosing between the different scenarios, we compute for each MCMC run the Bayesian Information Criterion (BIC, \citealt{Schwarz_1978}) and we select the run with the lowest value. The BIC is defined as
\begin{equation}
{\rm BIC} = k \ln{N} - 2 \ln{L(\hat{\theta})}
\label{eq:bic}
\end{equation}
where $k$ is the number of free parameters of each model, $N$ is the number of data points and $L(\hat{\theta})$ is the likelihood function of the model evaluated with the set of parameters $\hat{\theta}$ that maximize it. The likelihood is assumed to be gaussian for all the parameters considered in this model.

\subsection{Uniform density}
\label{sec:Uniform density}

We first consider an approach that does not require the presence of a low-density ISM cavity, in order to show that this configuration appears inadequate to fit the data.
In this scenario, the jet moves and then decelerates in a uniform medium and, hence, we fix $R_{\rm c} = 0$ pc and $\delta = 1$. The MCMC run provides a converged solution, and the fit results are presented in Table \ref{tab:fit_params_jets} and are shown in Figure \ref{fig:angsep_fit_uncertainties_nocavity}. In this case, the model cannot adequately reproduce the final part of the jet motion, and we obtain a BIC value of 120.4. While the results for most of the parameters are acceptable, the jet appears to require a huge amount of energy ($10^{48}\sim$10$^{49}$ erg) in order to travel $\sim$0.5 pc and sweep matter in a standard ISM density, therefore we deem this scenario to be unlikely. Moreover, with a uniform density we would expect the jets to continuously interact with the ISM and thus to be constantly detected for their whole motion, with a uniform decrease in radio flux density (unless the jets are highly relativistic), similarly to what was observed in \maxieight{} \citep{Bright}, instead of having a phase of non-detection followed by a re-brightening and deceleration, as in the case of \maxithirt{}.

\subsection{Motion in a low-density ISM cavity}
\label{sec:Motion in a low-density ISM cavity}

\begin{figure*}
\begin{center}
\includegraphics[width=\textwidth]{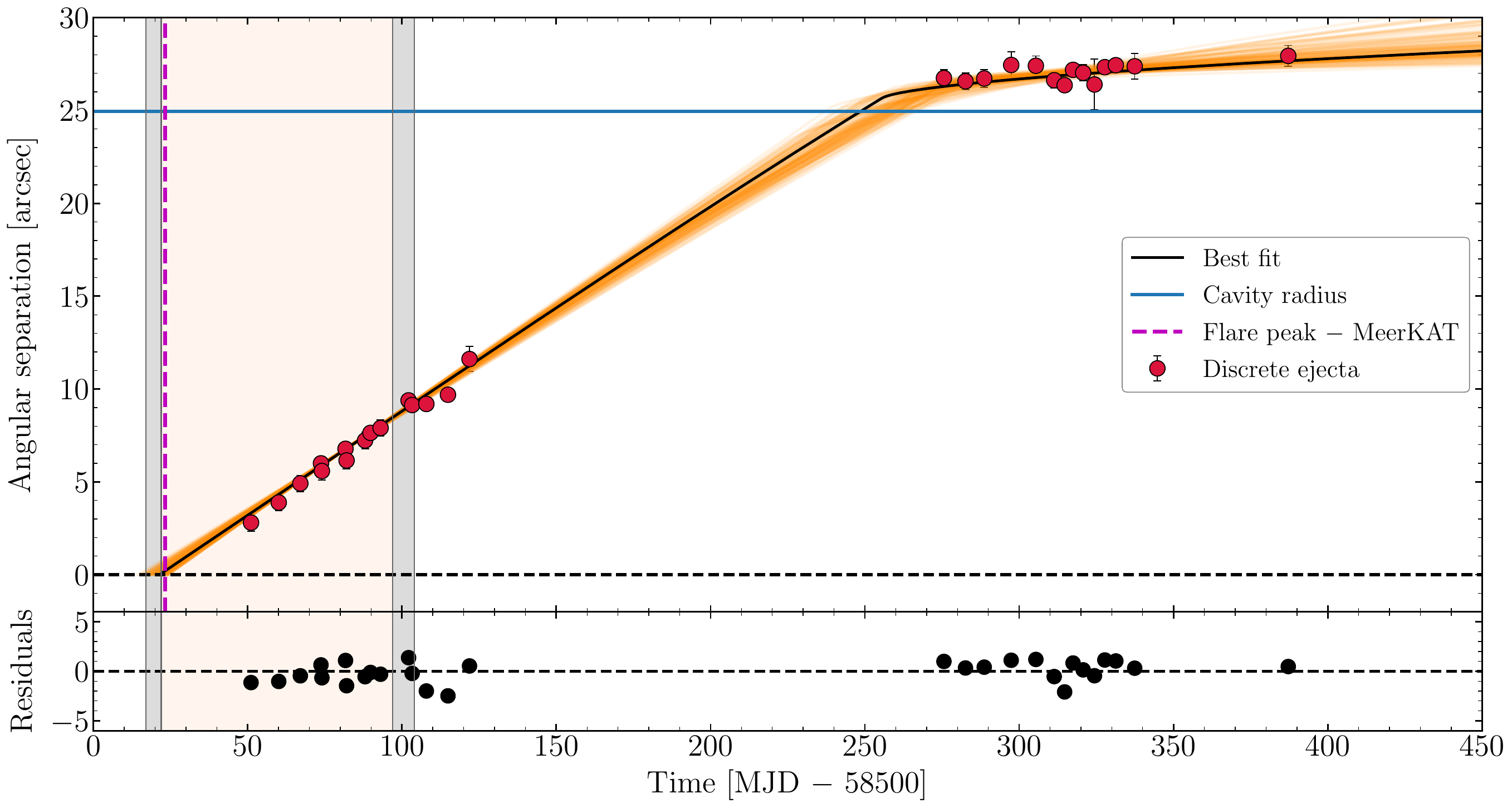}
\caption{Angular separation in arcsec between the discrete ejecta and the position of \maxithirt{}. We combine here MeerKAT 1.28 GHz and ATCA 5 GHz observations, shown as red points and taken from \protect\cite{Carotenuto2021}. The un-shaded, grey and seashell regions mark periods in which \maxithirt{} was, respectively, in the hard, intermediate and soft state \protect\citep{Zhang2020}. The black horizontal dashed line represents the zero separation from the core, while the time of the major flare observed with MeerKAT (MJD 58523) is represented with a magenta vertical line. The black continuous line represents the best fit obtained with the external shock model developed by \protect\cite{Wang_model} and the horizontal blue line shows the inferred radius of the low-density ISM cavity in which \maxithirt{} is embedded. The orange shaded area represents the total uncertainty on the fit and it is obtained by plotting the jet trajectories corresponding to the final positions of the MCMC walkers in the model parameter space. Residuals ([data –- model]/uncertainties) are reported in the bottom panel. The model appears to fit reasonably well our data, suggesting a jet deceleration due to the interaction with the wall of an ISM cavity.}
\label{fig:angsep_fit_uncertainties}
\end{center}
\end{figure*}

\begin{figure*}
\begin{center}
\includegraphics[width=\textwidth]{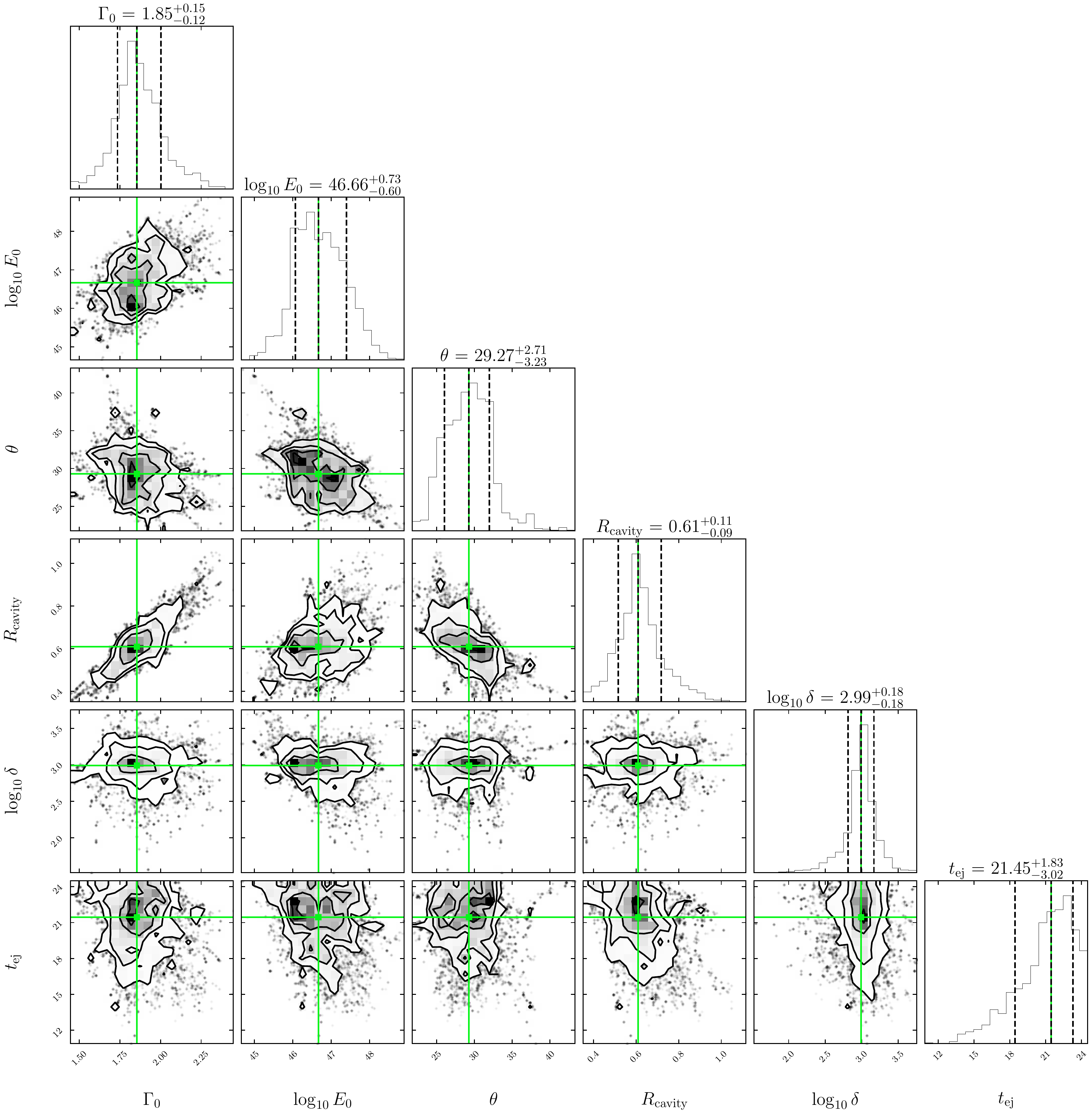}
\caption{Corner plots showing the results of the external shock model fit with MCMC. The panels on the diagonal show histograms of the one dimensional posterior distributions for the model parameters, including the jet initial Lorentz factor, initial energy, inclination angle and ejection time, as well as the radius and density jump of the low-density ISM cavity. The median value and the equivalent 1$\sigma$ uncertainty are marked with vertical dashed black lines.  The other panels show the 2-parameter correlations, with the best-fit values of the model parameters indicated by green lines/squares. The plot was made with the \textsc{corner} plotting package \citep{Foreman_mackey_2016}.}
\label{fig:corner}
\end{center}
\end{figure*}

Introducing a cavity and letting $R_{\rm c}$ and $\delta$ free to vary in the range of their priors results in a significantly improved fit.
The best fit results are reported in Table \ref{tab:fit_params_jets} and are shown in Figure \ref{fig:angsep_fit_uncertainties}, along with the proper motion of the jet. Uncertainties on the plot are represented as the trajectories corresponding to the final positions of the walkers in the parameter space.
In Figure \ref{fig:corner} we present the corner plot showing the one-dimensional posterior distribution for all the parameters and the 2-parameters correlations. 

The model appears to fit really well our data, and the good agreement with observations can be seen from the residuals in Figure \ref{fig:angsep_fit_uncertainties}. We obtain a BIC value of 49.5, which is significantly lower than the previous case, providing us with a strong evidence in favour redof the cavity scenario compared to a uniform density ISM. 

The motion that we obtain is similar to the \enquote{linear+Sedov} phenomenological model used in \cite{Carotenuto2021}, although in this case a physical model is used to explain the whole motion. According to this model, the jet is launched at $t_{\rm ej} = 21.5^{+1.8}_{-3.0}$ (MJD $- 58500$) with a bulk Lorentz factor $\Gamma_0 = 1.85^{+0.15}_{-0.12}$, corresponding to a mildly-relativistic speed, with an impressive amount of energy $E_0 = 4.6^{+20.0}_{-3.4} \times 10^{46}$ erg, and with a small inclination angle $\theta = 29.3\degree_{-3.2\degree}^{+2.7\degree}$. We infer that \maxithirt{} lives in a cavity with a radius $R_{\rm c} = 0.61^{+0.11}_{-0.09}$ pc and a large density jump $\delta = 980^{+514}_{-359}$.  In this context, our jet might have traveled undisturbed for the first phase, between MJD 58521 and 58750, adiabatically expanding and radiating in an environment with a low density of $n_{\rm ism}/\delta = 0.0010^{+0.0005}_{-0.0003}$ cm$^{-3}$, and then it decelerated and re-brightened after hitting the wall of the cavity, transitioning from a relativistic phase to a Sedov expanding phase around MJD 58750, with a sharp drop of $\Gamma$ from $\sim$1.8 to $\sim$1.
At the cavity wall, the jump in the ISM density caused the production of external shocks between the jet and the surrounding material, resulting in \textit{in-situ} particle acceleration responsible for the radio emission. This emission has been observed with \textit{Chandra} to extend up to the X-rays for some decelerating jets (e.g.\ \citealt{Corbel2002_xte, Corbel2005_h17, Espinasse_xray}). However, in the case of \maxithirt{} it has not been possible to observe the jets with \textit{Chandra}. This was due to the simultaneous brightness of the core of the system during the several reflares that followed the main outburst (see \citealt{Carotenuto2021}), since a bright X-ray core is expected to completely out-shine and engulf the resolved (but faint) X-ray emission from the ejecta.

\subsection{Impact of the dust-scattering halo distance}
\label{sec:Dust-scattering halo distance}

We note that \cite{Lamer_2021} recently reported a distance $D = 3.39 \pm 0.34$ kpc from observations of X-ray light echoes from a degree-scale dust scattering ring with eROSITA. While in this paper we center our prior on $D = 2.2$ kpc, we performed a MCMC run centering the distance prior on 3.39 kpc, with a sigma value of 0.4 kpc and with the same setup described in Section \ref{sec:Results} and \ref{sec:Motion in a low-density ISM cavity}. We obtain a good fit with a slightly higher BIC value of 63.3. The fit results are shown in Figure \ref{fig:angsep_fit_uncertainties_erosita}. As expected, a larger distance results in larger estimate for the jet parameters with respect to Section \ref{sec:Motion in a low-density ISM cavity}, of factors between 20\% and 50\%, with the exception of $E_0$, which has to be $\sim$3.5 times larger than what presented in the previous section in order to sustain a jet with the same proper motion at a distance $\sim$50\% larger.
While this distance might be plausible, the choice of the distance prior does not significantly affect the results and the main conclusions of this paper.

\section{Discussion}
\label{sec:Discussion}

We successfully modelled the motion of the large-scale jet from \maxithirt{} with a full dynamical model based on external shocks, adopting a Bayesian approach. The model appears to describe reasonably well the evolution of the jet in the different phases of its motion, and the fit we performed allows us to constrain with good or sufficient accuracy the physical parameters of the jet and of its environment. We discuss the obtained values for all the model parameters individually in the following sections, and stress that these results were achieved thanks to the dense and deep radio coverage of the jet deceleration phase.

\subsection{Initial Lorentz factor}
\label{sec:Initial Lorentz factor}

We obtain a precise estimate of the initial Lorentz factor of the jet: $\Gamma_0 = 1.85_{-0.12}^{+0.15}$, which implies a mildly relativistic ejecta, despite the high initial proper motion observed. By sampling from the best-fit posterior distribution of $\Gamma_0$, we can then infer the corresponding initial intrinsic jet speed $\beta_0 = (1-\Gamma_0^{-2})^{1/2} = 0.84^{+0.02}_{-0.03}$, which is consistent with the observational constraints presented in \cite{Carotenuto2021}.

It is difficult to constrain the Lorentz factor of jets from BH LMXBs by the observations of their proper motions, especially if the ejecta are significantly relativistic \citep{Fender_2003}, given that in those cases the source will be close to the maximum allowed distance $D_{\rm max} = c/\sqrt{\mu_{\rm app} \mu_{\rm rec}}$ \citep{Fender1999}, which can be obtained from simultaneous observations of the approaching and receding components. Therefore, for most of the sources only lower limits on $\Gamma$ can be placed (e.g.\ \citealt{Fender_2003, Miller-jones2006, Bright}). Jets from \maxithirt{} are, hence, likely to be less relativistic that the majority of bipolar ejecta from BH XRBs, and the inferred value of $\Gamma_0$ is broadly consistent with what observed for the less relativistic jets reported in \cite{Miller-jones2006}.
This is also supported by a rough estimation of $D_{\rm max}$, which can be obtained by solving for $D$ the proper motion equation \citep{Rees_1966, Mirabel_1992}
\begin{equation}
\mu_{\rm app} = \frac{\beta \sin\theta}{1 - \beta \cos{\theta}} \frac{c}{D}
\label{eq:proper_motion}
\end{equation}
assuming a maximum intrinsic speed $\beta = 1$ and the inferred values for $\theta$ and $\mu_{\rm app} \simeq 109$ mas day$^{-1}$. We obtain $D_{\rm max} \simeq 4$ kpc, which is significantly larger than the measured distance of \maxithirt{} (see Section \ref{sec:maxithirt}).

By applying the same external shock model, it was not possible to precisely constrain $\Gamma_0$ for the two-sided ejecta of \xte{} \citep{Steiner_xte} and \hh{} \citep{Steiner_h17}, but only to place lower limits on it. On the other hand, in our case we have a uniform and dense coverage of the first phase of the jet motion, which allows us to uniquely determine the jet intrinsic speed, given the reasonable results obtained for $\theta$ and $D$ (see Section \ref{sec:Inclination angle}).

\subsection{Inclination angle}
\label{sec:Inclination angle}

We obtain a medium-low inclination angle of the jet axis $\theta = 29.3\degree_{-3.2\degree}^{+2.7\degree}$, which suggests a relatively face-on system and it is consistent with the results presented in \cite{Carotenuto2021}. However, we note that the posterior distribution obtained for $\theta$ clearly traces its prior (see Figure \ref{fig:corner} and Table \ref{tab:fit_params_jets}), implying that this parameter is not completely constrained from the fit. The prior choice took into account the observational constraints from \cite{Carotenuto2021}. Nevertheless, the model reproduces the jet motion very well with such inclination angle, which is likely to be fairly close to the true value.

Interestingly, the inclination angle of the accretion disk was inferred to be $i = 28\degree \pm 3\degree$ from X-ray spectroscopy with \textit{NuSTAR}, using relativistic reflection models \citep{Anczarski_2020}. A measurement of the orientation of the orbital plane of the system would be extremely valuable to test the potential alignment between the disk and the jet axis. Since the jet axis is presumed to be aligned with the BH's spin axis, such measurement would be also important to test the alignment between the spin axis and the orbital plane. As already mentioned by \cite{Steiner_xte}, a good alignment between the two provides support for the BH spin estimation via the X-ray continuum-fitting methods \citep{Zhang_1997}. However, misaligned jets have been observed (e.g.\ \citealt{Miller-Jones2019, Poutanen_2021}), so such alignment is likely not to be universal among BH LMXBs.

With the values obtained for $\Gamma_0$ and $\theta$, we can compute the Doppler factor for the two ejecta. The Doppler factor for the approaching component at launch is $\delta_{\rm app} = \Gamma_0^{-1}(1 - \beta_0 \cos{\theta})^{-1} \simeq 2$, and such value remains stable for the first part of the jet motion, as inside the cavity $\Gamma \simeq \Gamma_0$. This implies that the received radio flux is boosted by a factor $\delta_{\rm app}^{3-\alpha} \simeq 16$, considering a discrete ejecta with a radio spectral index $\alpha \simeq -1$ \citep{Carotenuto2021}, greatly easing the detection the approaching ejecta, which peaked at a flux density of $\sim$6.7 mJy at 1.3 GHz \citep{Carotenuto2021}. For the un-detected receding component, the Doppler factor is $\delta_{\rm rec} = \Gamma^{-1}(1 + \beta \cos{\theta})^{-1} \simeq 0.3$, implying that the received flux is reduced by a factor $1/\delta_{\rm rec}^{3-\alpha} \simeq 125$, which explains why the receding component was never detected in any of our radio observations.

\subsection{Ejection date}
\label{sec:Ejection date}

The ejection date is one of the key parameters in the external shock model. From the fit, we infer the ejection date $t_{\rm ej} = 21.5^{+1.8}_{-3.0}$ (MJD $- 58500$), which is roughly two days before the peak of the $\sim$0.5 Jy radio flare observed with MeerKAT and, since this value is obtained with a full dynamical model, it is likely to be more reliable than the ejection date presented in \cite{Carotenuto2021}.
The estimation of the time of ejection is fundamental to reconstruct the precise evolution of the system during state transitions and to understand the contribution of the various physical components of the system (e.g.\ compact jets, corona, inner accretion flow) in the production and acceleration of the discrete ejecta. The jet ejection is inferred to happen while \maxithirt{} was in the HIMS, 
roughly one day before the first detection of Type-B QPOs with NICER on MJD 58522.6 \citep{Zhang2020, Zhang_2021}. Similarly to the conclusions of \cite{Carotenuto2021}, in our case Type-B QPOs appeared to be produced after the jet ejection, as already observed for \hh{} and \maxififth{} \citep{Miller-Jones_h1743, Russell_1535}, and were not simultaneous with the jet launch, as instead found for \maxieight{} \citep{Bright, Homan_qpo}, hence the potential causal link between the two phenomena remains unclear. To investigate such sequence of events more in depth, the NICER timing results on \maxithirt{} will be put in relation to the inferred ejection date in a future work.

\subsection{A large kinetic energy}
\label{sec:A large kinetic energy}

The ejecta is inferred to have an extremely large kinetic energy at the time of ejection: as one of the key outputs of the model, we obtain
$E_0 = 4.6^{+20.0}_{-3.4} \times 10^{46}$ erg.
The result is particularly interesting because such energy estimation is several orders of magnitude larger than what was estimated for \maxithirt{} ($\sim$10$^{42}$ erg; \citealt{Carotenuto2021}) and from what is generally inferred for ejecta from XRBs (see for instance \citealt{Fender1999, Tetarenko2017, Rushton, Russell_1535, Miller-Jones2019, Espinasse_xray}), which is obtained by considering the minimum energy required to produce the observed radio synchrotron radiation, assuming equipartition between particles and magnetic fields (e.g.\ \citealt{Longair}). Our estimation is instead consistent with the kinetic energy obtained for the first jets discovered in \grs{} \citep{Mirabel1994}, and in \xte{} (by applying the same external shock model; \citealt{Steiner_xte}), and it is slightly higher than what obtained for \hh{} \citep{Steiner_h17}, whose jets decelerate much earlier compared to \maxithirt{}. These findings support the recent claim that the total energy of these jets is not well-traced by the observed electromagnetic emission, and this leads to a significant underestimation of the energetic content of these objects (e.g.\ \citealt{Bright}). As a result, it is likely that jets only radiate away a small fraction of their total energy, which is instead almost completely transferred to the environment, as in the case of \maxieight{} \citep{Bright}. This implies that the overall feedback of the jet on the surrounding environment is possibly way larger than previously thought, and, at the same time, it justifies our assumption in the model of negligible radiative losses and consequent adiabatic expansion of the jet.

There is, however, a \textit{caveat} on such energy estimation due to the degeneracy between $E_0$, $n_{\rm ism}$ and $\phi$, since the model can only constrain the value $E_0/n_{\rm ism}\phi^2$, and, hence with our assumptions we essentially obtain
\begin{equation}
E_0 = 4.6^{+20.0}_{-3.4} \times 10^{46} \ \left(\frac{n_{\rm ism}} {1 \ {\rm cm^{-3}}}\right) \left(\frac{\phi}{1\degree}\right)^2 \ {\rm erg.}
\label{eq:energy_result}
\end{equation}
In Section \ref{sec:The external shock model} we discussed and motivated our choice of assuming the reasonable values of $n_{\rm ism} = 1$ cm$^{-3}$ and $\phi = 1\degree$. The energy could be overestimated in case the opening angle is much smaller than the $1\degree$ assumed. However, if we imagine our opening angle to be $\sim 10\%$ of that value, $E_0$ would be $\sim$1\% of what reported above, but it would still be roughly two orders of magnitude larger than the radiated energy. Moreover, while for the majority of XRB jets we have un-constraining upper limits (larger than 1\degree) on their opening angle \citep{Miller-jones2006}, such ejecta have been in some cases resolved with VLBI or in X-rays (e.g.\ \citealt{Miller-Jones_2004, Rushton, Espinasse_xray}), yielding half-opening angles larger than 1\degree, hence it appears to be unlikely that jets in \maxithirt{} could have significant smaller apertures. In addition, it is difficult to imagine how such jets could be confined in such low-density environments surrounding microquasars \citep{Heinz_2002, Hao, Miller-jones2006}.
Another possibility is that the density of the external ISM, \textit{outside} the cavity, is smaller than what we assumed for $n_{\rm ism}$, meaning that \maxithirt{} could be located in a region with a particular hot phase of the ISM. However, this is unlikely given that our source lies in the galactic plane, where we expect higher ISM densities (e.g.\ \citealt{Cox_2005}). Therefore, our estimate of $E_0$ appears to be reliable, and the evidence for a highly energetic jet appears to be robust.

\subsection{Jet power}
\label{sec:Jet power}

The energy $E_0$ discussed in Section \ref{sec:A large kinetic energy} must be provided by the system in a time $\Delta t$, and hence we can consider now the average power required to accelerate the jets in \maxithirt{}, which is $P_{\rm jet} = 2E_0/\Delta t$ (for a bipolar ejection).
The exact value of $\Delta t$ is unknown, but it is possible to obtain a zero-order estimate by considering the interval between our inferred $t_{\rm ej}$ at which the jet was launched and the peaking time of the radio flare produced by the jet on MJD 58523.22 observed with MeerKAT \citep{Carotenuto2021}. We obtain an observed ejection timescale of $\sim$1.8 days, which, given the relativistic motion of the jet, has to be corrected to the source rest frame through $\Delta t_{\rm obs} = \delta_{\rm app}^{-1} \Delta t_{\rm RF} $. We obtain a rest-frame rising timescale of $\sim$3.6 days, which leads to a total mechanical power of the jets
\begin{equation}
P_{\rm jet} \simeq 3 \times 10^{41} \ \left(\frac{n_{\rm ism}} {1 \ {\rm cm^{-3}}}\right) \left(\frac{\phi}{1^o}\right)^2 \left( \frac{\Delta t_{\rm RF}}{3.6 \ {\rm d}} \right)^{-1} \ {\rm erg \ s^{-1}}.
\label{eq:jet_power}
\end{equation}
As can be expected by the large value of $E_0$ and the short ejection timescale, such power is extremely high, and it has to be compared with the simultaneously available accretion power. This can be roughly estimated from the simultaneous bolometric X-ray luminosity, since $L_{\rm X} = \eta \dot{M}c^2$, where $\eta$ is the radiative efficiency of the accretion flow. $L_{\rm X}$ can be obtained from the 1--10 keV X-ray flux observed with \textit{Swift} on MJD 58521 \citep{Carotenuto2021}, where the bolometric luminosity is estimated using the conversion factor from \cite{Migliari_fender_2006}. This estimation yields an X-ray luminosity $\sim 7.9 \times 10^{38}$ erg s$^{-1}$. Such value is $\sim$61\% of the Eddington luminosity, which is standard limit for stable accretion ($\simeq 1.3 \times 10^{39}$ erg s$^{-1}$ for a $10 M_{\odot}$ BH accreting hydrogen).
With the standard assumption $\eta = 0.1$, we obtain an accretion power of $\sim 7.9 \times 10^{39}$ erg s$^{-1}$. Therefore, our inferred value of $P_{\rm jet}$ is be $\sim$40 times higher than the available accretion power, implying a discrepancy between the available accretion power and the power required by the jets. Such value of $P_{\rm jet}/\dot{M}c^2$ (which can define as $\eta_{\rm jet}$) appears to be too large even if we include the BH spin as an additional energy reservoir, that can be extracted from the compact object through the Blandford-Znajek mechanism \citep{Blandford_1977}. Simulations show that $\eta_{\rm jet}$ can attain the maximum value of $\sim$1.4 for jets produced by rapidly spinning BHs \citep{Tchekhovskoy_2011}, well below our estimate.

However, both the quantities that we are comparing, $P_{\rm jet}$ and $\dot{M}c^2$, are affected by uncertainties in their estimation, and the tension between the two can possibly be mitigated in several ways. Since $P_{\rm jet}$
depends on the assumed values of $n_{\rm ism}$ and $\phi$, it is possible that such quantities are in fact smaller than the ones we assumed (see Section \ref{sec:A large kinetic energy}). Hence, we can place upper limits on the two parameters that cannot be constrained from the model alone. Starting from $P_{\rm jet} \leq 1.4 \dot{M}c^2$, if we assume that $n_{\rm ism} = 1$ cm$^{-3}$, then we have that $\phi \leq 0.2\degree$. On the other hand, by assuming $\phi \leq 1\degree$, we can infer that $n_{\rm ism} \leq 0.04$ cm$^{-3}$. However, while in principle these values are possible, both the upper limits appear to be unlikely: an unusually narrow jet seems difficult to understand from the arguments discussed in Section \ref{sec:A large kinetic energy}, while a low value of the ISM density outside a cavity appears not plausible as well, given the location of \maxithirt{} in the galactic plane. Moreover, a lower value of $n_{\rm ism}$ would imply an even smaller value $n_{\rm ism}/\delta$ of the density inside the cavity, which is hard to justify.

Another reason for such discrepancy could lie in an incorrect estimation of the ejection timescale, as a larger $\Delta t$ would decrease our inferred value of $P_{\rm jet}$. While the uncertainty on the flare rising timescale is of the order of days, and thus that value is unlikely to be heavily underestimated\footnote{Typical values for the radio flare rising timescales are of the order of hours to days (e.g.\ \citealt{Russell_1535, Bright, Homan_qpo}).}, it is possible that such timescale is not a good proxy for the timescale in which the system effectively accelerates the ejecta.

Finally, the accretion power could be higher than what we have estimated. The accretion efficiency $\eta$ might be lower than 0.1 (even if that is unlikely, given that the accretion flow should be radiatively efficient in the intermediate and soft state, e.g.\ \citealt{Done_2007}), or the X-ray luminosity might have been underestimated. While our \textit{Swift} measurement of the X-ray luminosity is not exactly simultaneous as our inferred ejection date, we speculate that in this case the main source of uncertainty lies in the conversion from the luminosity in the 1-10 keV energy range to the bolometric X-ray luminosity $L_{\rm X}$. A larger $L_{\rm X}$, possibly above the Eddington limit for a limited period of time, might be a plausible explanation for such a large discrepancy. Short super-Eddington phases linked to major ejections have likely taken place in other systems, as for instance \grs{} \citep{Mirabel1994}, \xte{} \citep{Steiner_xte} and possibly also \maxieight{} \citep{Bright}.
Persisting super-Eddington luminosities are usually observed in ultraluminous X-ray sources (ULXs), a part of which is composed of neutron stars accreting above the Eddington limit (e.g.\ \citealt{Kaaret_2017}).

We speculate that a combination of the options outlined above could explain the discrepancy between our inferred jet power and the available accretion power. We remark that obtaining a reliable estimate of the jet power is of prime importance for understanding of jet production and acceleration mechanisms.

\subsection{Jet mass}
\label{sec:jet_mass}

\begin{figure}
\begin{center}
\includegraphics[width=\columnwidth]{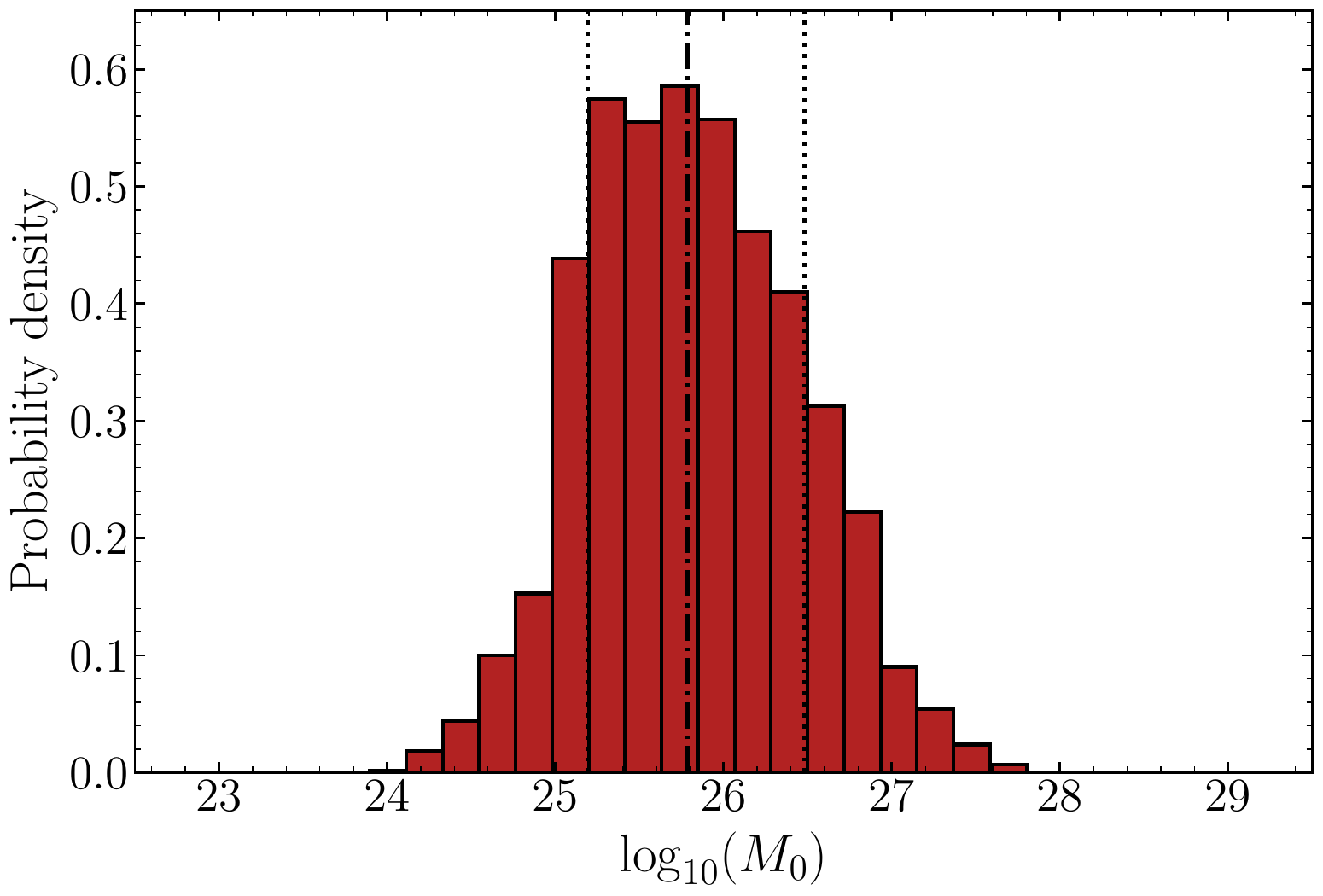}
\caption{Posterior distribution for the ejecta mass $M_0$ in grams, obtained using the chain samples produced for $E_0$ and $\Gamma_0$. The median value and the equivalent 1$\sigma$ uncertainty are marked, respectively, with vertical dot-dashed and dotted black lines. Assuming $n_{\rm ism} = 1$ cm$^{-3}$ and $\phi = 1\degree$, we infer a jet mass $M_0 = 6.1_{-4.5}^{+24.1} \times 10^{25}$ g.}
\label{fig:mass}
\end{center}
\end{figure}

From Equation \ref{eq:energy}, we are able to constrain the mass of the ejecta $M_0$ by sampling from the posterior distributions of  $E_0$ and $\Gamma_0$ obtained with the MCMC fit. The resulting posterior distribution for $M_0$ is shown in Figure \ref{fig:mass}. We infer a jet mass
\begin{equation}
M_0 = 6.1_{-4.5}^{+24.1} \times 10^{25}    \left(\frac{n_{\rm ism}} {1 \ {\rm cm^{-3}}}\right) \left(\frac{\phi}{1^o}\right)^2 {\rm g,}
\end{equation}
corresponding to $\sim 3.1 \times 10^{-8} M_{\odot}$. While this estimation is consistent with the first jets discovered in \grs{} \citep{Mirabel1994}, obtained assuming one proton per electron, it is larger than the estimated masses for other ejecta observed in the same source \citep{Mirabel_1998, Fender1999}, and in other sources (e.g.\ \citealt{Steiner_xte, Espinasse_xray}). It appears that, as many other physical parameters, the mass content of the jets can greatly vary among the population of XRBs, and significant variations are observed even between multiple ejecta from the same source on short timescales \citep{Tetarenko2017}.
However, the mass estimations for discrete ejecta are very uncertain, and often depend on parameters that are poorly constrained, such as the jet composition (specifically the proton content) and the Lorentz factor. In our case, we have also to consider the additional dependence on the two parameters $n_{\rm ism}$ and $\phi$, but for which plausible values were assumed (see Section \ref{sec:A large kinetic energy}). Therefore, more observations are needed to increase the statistics and reduce the uncertainties on these physical quantities.

By considering the ejection timescale $\Delta t$ introduced in Section \ref{sec:Jet power}, we can estimate the mass outflow rate to be $\simeq 3.9 \times 10^{20}$ g s$^{-1}$ (for a bipolar ejection), implying a huge transfer of mass from the system to the jet during the ejection. Not surprisingly, such outflow rate is higher than what was obtained for other sources (e.g. SS~433 and \grs{}, \ \citealt{Kotani_1996, Fender1999}), and higher than the accretion rate of a system radiating at a significant fraction of the Eddington luminosity. This possibly implies that during the formation and launch of discrete ejecta, the majority of the inflowing mass is directly channeled into the jets and it is not advected into the event horizon \citep{Fender_1998,Fender1999}. Transitory mass outflow rates above the Eddington limit on the accretion rate are also consistent with simulations of accretion disks of stellar mass BHs \citep{Okuda_2005}. Such mass outflow rate is also higher than the mass loss rate commonly attributed to disk winds, that can be as large as roughly ten times the mass accretion rate \citep{Ponti_2012}.
However, disk winds have not been detected for \maxithirt{} to our knowledge, hence a direct comparison of the two mass outflow rates is not possible.

\subsection{Jet composition}
\label{sec:Jet composition}

The composition of discrete ejecta is generally unknown. Evidences for a baryonic content have been found in the jets from SS~433 (possibly from the ISM matter entrained by the jets), with the detection of Doppler-shifted iron line emission \citep{Kotani_1996, Migliari_2002}, but the rest of known ejecta displays featureless synchrotron spectra and might be composed of proton-electron or electron-positron plasma \citep{Fender2006}.

Hints on the composition of the ejecta from \maxithirt{} can be obtained by estimating the number of protons and electrons required to match the observational constraints from the jet radiation and motion. To begin, the number of relativistic electrons in the jet can be obtained through widely-applied minimum energy calculations \citep{Longair}, assuming a standard electron energy distribution index $p = 2$, which corresponds to a radio spectral index $\alpha = -0.5$, and integrating it between the minimum and maximum energies of the electron population. The electron number $N_{\rm e}$ is then proportional to
\begin{equation}
N_{\rm e} \propto \nu^{\frac{p-1}{2}}  \left[ \frac{L_{\nu}}{A(\alpha) B}    \right] \left[\nu_{\rm min}^{\frac{1-p}{2}}-\nu_{\rm max}^{\frac{1-p}{2}} \right] 
\label{eq:electron_number}
\end{equation}
where the full details of the calculations are presented in Appendix \ref{sec:number of electrons}. Here $L_{\nu} = 4\pi D^2 S_{\nu}$ is the monochromatic luminosity of the synchrotron flare, with a flux density $S_{\nu} \simeq 486$ mJy observed at a frequency $\nu = 1.28$ GHz with MeerKAT on MJD 58523 \citep{Carotenuto2021}. We assume that the electrons radiate between a minimum frequency $\nu_{\rm min} = 10^9$ Hz (lower limit of the MeerKAT observing band) and a maximum frequency $\nu_{\rm max} = 10^{15}$ Hz. The real value of both frequencies is unknown, but the results depend very weakly on $\nu_{\rm max}$, since the emission is dominated by the low-energy electrons. Moreover, $A(\alpha)$ is a constant (see Section \ref{sec:number of electrons}) and $B$ is the magnetic field at equipartition, which can be written as \citep{Longair}
\begin{equation}
B = \left[ \frac{3\mu_0}{2}   \frac{G(\alpha)\hat{\eta} L_{\nu}}{V}   \right]^{\frac{2}{7}}
\label{eq:B_field}
\end{equation}
where $\mu_0$ is the permeability of free space, $G(\alpha)$ is a constant discussed in Section \ref{sec:number of electrons} that depends weakly on $\alpha$, $\nu_{\rm min}$ and $\nu_{\rm max}$, $V$ is the size of the emitting region and $\hat{\eta}$ is a parameter linked to the ratio of energy in protons to that in electrons, and we assume for it the standard value of 1. In particular, the volume $V$ of the jet at launch (when it produced the synchrotron flare) is a key parameter for the calculation of $N_{\rm e}$, and we estimate it by considering the ejection timescale $\Delta t$, discussed in Section \ref{sec:Jet power}, and a reasonable jet expansion speed of $0.05 c$, which is consistent with results from \cite{Fender_2019_equipartition} and with observational constraints from \cite{Carotenuto2021}.
We take into account Doppler boosting due to the relativistic speed of the jet in the first phase of its motion, and we convert our measured quantities to the jet rest frame by correcting the observed frequency with $\nu_{\rm obs} = \delta_{\rm app} \nu_{\rm rest \ frame}$, the observed ejection timescale $\Delta t_{\rm obs} = \delta_{\rm app}^{-1} \Delta t_{\rm rest \ frame}$
and observed flux $S_{\rm \nu,obs} = \delta_{\rm app}^{3-\alpha} S_{\rm \nu,rest \ frame}$, assuming $\alpha = -0.5$ from \cite{Carotenuto2021}.
An upper limit on the size of the emitting region can also be obtained by considering the unresolved radio point source and the MeerKAT beam on MJD 58523, but this would lead to an overestimation of the actual volume $V$ due the $\sim$5 arcsec MeerKAT L-band resolution.
We obtain $N_{\rm e} \simeq 3 \times 10^{43}$, which is smaller compared to other discrete ejecta (e.g.\ \citealt{Fender1999, Espinasse_xray}). If we consider that each electron is accompanied by a cold proton, we obtain a total jet mass $N_{\rm e}(m_{\rm e} + m_{\rm p}) \simeq 5 \times 10^{19}$ g, which is much smaller than our inferred total jet rest mass, discussed in Section \ref{sec:jet_mass}. Therefore, there seems to be evidence for the jet to be made up of non-radiating particles. Such jet could be possibly dominated by \enquote{cold} protons which would carry most of the kinetic energy. In principle, the number of protons could simply be estimated as
\begin{equation}
N_{\rm p} = \frac{M_0}{m_{\rm p}} \simeq 4 \times 10^{49} \left(\frac{n_{\rm ism}} {1 \ {\rm cm^{-3}}}\right) \left(\frac{\phi}{1^o}\right)^2 
\label{eq:proton_number}
\end{equation}
which greatly exceeds the number of synchrotron-emitting relativistic electrons, even taking into account the uncertainty due to the dependence on $n_{\rm ism}$ and $\phi$ (see Section \ref{sec:A large kinetic energy}). We could speculate that the jet might include a population of non-relativistic (and hence non-radiating) electrons coupled to the cold protons for charge balance. However, such estimation is purely qualitative, and this is due to the large uncertainty present on many parameters involved in this calculation, and to our likely over-simplification of the problem.

The discrepancy between $N_{\rm e}$ and $M_0$ could be alleviated in several ways: with a more precise estimation of the jet volume at launch (which could be underestimated), or if the jet plasma is far from equipartition, with a stronger magnetic field. Moreover, electrons in the jet could radiate even below $\nu_{\rm min}$, with more protons associated, although there is not yet observational evidence of ejecta radiating at lower radio frequencies, as argued in \cite{Fender1999}. Lastly, $E_0$, and consequently $M_0$, might be overestimated, in particular if the ejecta has a significantly smaller opening angle $\phi$ than what assumed in this work, although we deem this as unlikely, as discussed in Section \ref{sec:A large kinetic energy}.

\subsection{The ISM cavity}
\label{sec:The ISM cavity}

The external shock model applied in this work considers the source to be embedded in a low density cavity present in the ISM. After a first phase of constant speed motion, the jet hits the cavity border and decelerates as a result of the interaction with a much denser environment. From the fit, we infer the low-density ISM cavity to have a radius $R_{\rm c} = 0.61^{+0.11}_{-0.09}$ pc, which is consistent with the constraints from \cite{Carotenuto2021}.
We also infer a density jump $\delta = 980^{+514}_{-359}$. As mentioned in Section \ref{sec:Motion in a low-density ISM cavity}, this implies a density $ n_{\rm ism}/\delta = 0.0010^{+0.0005}_{-0.0003}$ cm$^{-3}$ inside the empty bubble, assuming a standard external ISM external density $n_{\rm ism} = 1$ cm$^{-3}$. Interestingly, the value of $\delta$ is roughly one order of magnitude larger than the ones inferred for other BH XRBs possibly located inside a cavity, such as \xte{} and \hh{} \citep{Wang_model, Hao, Steiner_xte, Migliori2017}.
However, our density estimate is consistent with what was inferred for the environments of \grs{} and GRO~J1655--40, obtained assuming that the discrete ejecta are travelling in a constant density medium up to distances of at least $\sim$0.04 pc \citep{Heinz_2002}.
The size of the cavity $R_{\rm c}$ is consistent with what obtained for \xte{} and \hh{} (e.g.\ \citealt{Hao}). XTE~J1752--223 might also be located in an under-dense cavity, although such bubble would have a smaller size compared to other sources \citep{Yang2010, Miller_jones_sedov}.

As argued in \cite{Carotenuto2021}, \maxithirt{} is fully consistent with the scenario of microquasars embedded in low density environments proposed by \cite{Hao}. It is currently unclear how such cavities might be produced. BH XRBs might be located in small regions occupied by the hot ISM phase, or, more likely, the low-density region could result from the system's activity \citep{Heinz_2002, Hao}.
The stability of these cavity would depend on the pressure balance with the external ISM, and their evolution is likely to take place on timescales much longer than the common recurrence outburst times for XRBs, which is estimated to span from years to centuries \citep{Remillard_xrb, Corral_santana}. Even if we imagine the cavity to be completely empty, it would take roughly $\sim 5 \times 10^4$ years (much longer then typical XRB evolving timescale) for the external gas expanding at the sound speed to refill a $\sim$0.6 pc cavity, assuming an external ISM with density $n_{\rm ism} = 1$ cm$^{-3}$ and temperature $\sim$8000 K \citep{Ferriere_2001}.

An empty bubble might be created by the supernova that produced the BH in \maxithirt{} in case of low kick-out velocity, although we deem this first option as unlikely due to the fact that the system has to be significantly young for the ISM cavity not be yet refilled by the surrounding material. However, in that case we would expect to clearly detect the radio supernova remnant, for which there is no evidence \citep{Carotenuto2021}, suggesting that \maxithirt{} might be an old system. Several other possibilities involve previous outbursts and the feedback of jets or winds on the ISM surrounding the system.

The cavity might have been carved out by jet activity in previous outbursts. There are examples of jets from XRBs developing at large scales and creating hot spots at the terminal shock surface, where kinetic energy is continuously transferred from the system to the ISM \citep{Seward_1980, Mirabel_1992, Heinz_2007, GRS1758_nature, Coriat_2019}. A pc-scale cavity is also believed to be produced by the jets of Cygnus X--1 \citep{Gallo_2005}. In this case, and in absence of fast jet precession, we would expect rather a tunnel or a narrow under-dense region instead of a cavity, which would justify the jet collimation but it would require frequent ejections from the system in order to be sustained over time, as argued in \cite{Hao}. A wide, symmetrical cavity might be created by more steady and uncollimated outflows, such as winds, which could be produced by the companion star (e.g.\ \citealt{Sell_2015}), by the more massive progenitor of the compact object (e.g.\ \citealt{Gaensler_2005}), or by the accretion disk itself during phases of outburst (e.g.\ \citealt{Miller_2006, Fuchs_2006, Munoz_darias_2019}). While the companion stars of BH LMXBs might not be massive enough to produce strong winds, the accretion disks represent more plausible candidates, possibly displaying high mass outflow rates (e.g.\ \citealt{Munoz_darias_2016}). Due to their non-relativistic speeds, winds have necessarily to be produced during previous outbursts in order to push away the ISM and produce the cavity, since the mildly-relativistic ejecta from \maxithirt{} is way faster than the winds and it is ejected close to the beginning of the outburst.
However, it is not clear at this stage if the winds have enough power to carve out cavities at at pc-scale \citep{Hao}. 

\subsubsection{Alternative explanations}
\label{subsec:Alternative explanations}

A scenario alternative to the cavity might involve a denser ISM region, such as a molecular cloud, encountered by the jet on its path, which could be responsible for the sudden deceleration of the ejecta. However, this is unlikely because it would imply that the jet is travelling in an environment with a standard ISM density $n_{\rm ism} = 1$ cm$^{-3}$, before hitting a higher density region. From our results, such scenario would require an initial energy of the jet to be a factor $\sim$1000 larger than the estimated $E_0$ (see Sections \ref{sec:Motion in a low-density ISM cavity} and \ref{sec:A large kinetic energy}) to match the observations. This is due to the fact that the swept-up mass would be much higher than what expected in the cavity scenario, unless the jet is strongly collimated, with a half-opening angle $\phi \ll 1\degree$.

It is important to mention that a physical obstacle, such as the border of a cavity, is not strictly required for a relativistic object to produce a proper motion curve similar to the one of \maxithirt{}, shown in Figure \ref{fig:angsep_fit_uncertainties}. In fact, the object might have a high initial Lorentz factor and could be continuously decelerating in a uniform medium with constant density, as for instance could be the case in \maxieight{} \citep{Bright, Espinasse_xray}. The observed sudden deceleration would be simply the signature that the jet's Lorentz factor reached values $\Gamma \lesssim 2$, starting from values much higher than the ones obtained for our run of the model in a uniform density environment (Section \ref{sec:Uniform density}). In this context, Doppler boosting might explain the light curve of the highly relativistic jet, including the late-time re-brightening. Such interpretation is discussed in Fender \& Rhodes (\textit{in prep.}).

\subsection{The receding component}
\label{sec:The receding component}

The receding component paired to the discrete ejection considered in this work has never been detected in radio \citep{Carotenuto2021}, and Doppler de-boosting is likely responsible for this, as discussed in Section \ref{sec:Inclination angle}.
Assuming a perfectly bipolar ejection, and the same physical parameters obtained for the approaching component, we expect the receding component to display a proper motion $\mu_{\rm rec} = \beta c \sin{\theta} [D(1 + \beta\cos{\theta})]^{-1} \simeq 16.7$ mas day$^{-1}$. Hence, from the ejection time, it would take approximately 1500 days ($\sim$4.1 years) for the receding ejecta to reach the cavity border at 25 arcsec from the core position, assuming a symmetrical cavity (which appears to not be the case for \xte{}; \citealt{Hao, Steiner_xte}). This means that in our reference frame the receding jet has not reached yet the cavity border. It might be possible to detect emission from the receding component once the border is reached, since the ejecta will transition to a non-relativistic motion, synchrotron emission from re-accelerated particles will be produced and Doppler boosting will cease to be effective.

\section{Conclusions}
\label{sec:Conclusions}

In this paper, we have presented the application of the external shock model developed by \cite{Wang_model} to the decelerating discrete ejecta detected during the 2019/2020 outburst of \maxithirt{}. We fitted the jet angular distance data with a Bayesian approach and we found that the model provides an excellent description of the jet motion, from the first phase of high, constant speed to the last deceleration phase.
From the fit, we are able to obtain important insights on the physical parameters of the jet and of the system's environment.
We infer a mildly-relativistic jet with an initial Lorentz factor $\Gamma_0 = 1.85^{+0.15}_{-0.12}$ and with a low inclination angle with respect to the line of sight $\theta = 29.3\degree_{-3.2\degree}^{+2.7\degree}$, implying that Doppler boosting is likely responsible for the non-detection of the receding jet component.
The initial energy of the jet ($E_0 = 4.6^{+20.0}_{-3.4} \times 10^{46}$ erg) is very large and it provides support to the recent claim that ejecta from BH LMXBs do not radiate away most of their energy, which is instead largely transferred to the surrounding environment. Due to the large amount of energy contained in the jet and the short launching timescale, its ejection presumably required a power larger than what is available from accretion only, and we discuss in this work several options that might cause this discrepancy, including the possible underestimation of the accretion power and/or of the launching timescale, or the overestimation of two key parameters, $n_{\rm ism}$ and $\phi$, that cannot be constrained from the model alone. Additional observations of discrete ejecta from BH LMXBs, with detailed estimations of the jet energy and ejection time, will also be needed to improve our estimations of the true jet power.

We are able to place constraints on the jet mass and on its ejection date, and we infer that the matter content of such jet might be dominated by cold protons, which largely exceed the number of relativistic electrons responsible for the jet synchrotron emission.
The jet travels inside a low-density cavity which is inferred to have a radius $R_{\rm c} = 0.61^{+0.11}_{-0.09}$ pc and a density jump $\delta = 980^{+514}_{-359}$ with respect to the external standard ISM density $n_{\rm ism} = 1$ cm$^{-3}$, before encountering the cavity border and strongly decelerating, with possible \textit{in-situ} particle acceleration. 
This is the first time that such external shock model, which is simple but rich of physical information, is applied to a one-sided jet component. In this context, the coverage of the jet deceleration, which is not common among jets from XRBs, is fundamental, as it allows us to properly constrain the jet trajectory and, as a consequence, to constrain the model parameters. More observations of decelerating ejecta from BH LMXBs are needed in order to confirm our results and to increase the size of our sample, which will lead us to a significant improvement of the current understanding of the jet production, acceleration and feedback on the surrounding environment. In order to cover the whole jet motion, dense, uniform and sensitive radio monitoring campaigns are required, and the new generation of radio-interferometers, such as MeerKAT (which already detected a significant number of discrete ejecta), its upcoming upgraded version, SKA-MID \citep{Braun_2015} and the ngVLA \citep{Selina_2018} will be ideal to achieve this goal.

\section*{Data availability}
The un-calibrated MeerKAT and ATCA visibility data are publicly available at the SARAO and ATNF archives, respectively at \url{https://archive.sarao.ac.za} and \url{https://atoa.atnf.csiro.au}. The angular separation data used in this work are presented in Table \ref{tab:first_jet_angsep} and in \cite{Carotenuto2021}.

\section*{Acknowledgements}
We thank Jonathan Ferreira and Rob Fender for the useful discussions.
FC acknowledges support from the project Initiative d’Excellence (IdEx) of Universit\'{e} de Paris (ANR-18-IDEX-0001).
We acknowledge the use of the Nan\c cay Data Center, hosted by the Nan\c cay Radio Observatory (Observatoire de Paris-PSL, CNRS, Universit\'{e} d'Orl\'{e}ans), and also supported by Region Centre-Val de Loire.
Support for this work was provided by NASA through the NASA Hubble Fellowship grant \#HST-HF2-51494.001 awarded by the Space Telescope Science Institute, which is operated by the Association of Universities for Research in Astronomy, Inc., for NASA, under contract NAS5-26555.
This project also made use of \textsc{matplotlib} \citep{matplotlib}, \textsc{numpy} \citep{harris2020array} and Overleaf (\url{http://www.overleaf.com}).

%%%%%%%%%%%%%%%%%%%%%%%%%%%%%%%%%%%%%%%%%%%%%%%%%%

%%%%%%%%%%%%%%%%%%%% REFERENCES %%%%%%%%%%%%%%%%%%

% The best way to enter references is to use BibTeX:

\bibliographystyle{mnras}
\bibliography{maxi1348_mcmc_paper} 
% if your bibtex file is called example.bib

% Alternatively you could enter them by hand, like this:
% This method is tedious and prone to error if you have lots of references
%\begin{thebibliography}{99}
%\bibitem[\protect\citeauthoryear{Author}{2012}]{Author2012}
%Author A. N., 2013, Journal of Improbable Astronomy, 1, 1
%\bibitem[\protect\citeauthoryear{Others}{2013}]{Others2013}
%Others S., 2012, Journal of Interesting Stuff, 17, 198
%\end{thebibliography}

%%%%%%%%%%%%%%%%%%%%%%%%%%%%%%%%%%%%%%%%%%%%%%%%%%

%%%%%%%%%%%%%%%%% APPENDICES %%%%%%%%%%%%%%%%%%%%%

\appendix

\onecolumn

\section{Angular separation data}
\label{sec:angsep data}

\begin{ThreePartTable}

\begin{TableNotes}
  \item[a] {\footnotesize Combination of the MJD 58830 and 58831 ATCA epochs to obtain a higher significance detection of RK1 at both 5.5 and 9 GHz.}
\end{TableNotes}
\LTcapwidth=\textwidth
\begin{center}
\setlength{\extrarowheight}{.3em}
\begin{longtable}{*{5}{c}}
\caption{Measured positions of RK1 and its angular separation from \maxithirt{}, taken from \protect\cite{Carotenuto2021}. The errors reported for the coordinates are only the statistical ones from source fitting, while the error on the angular separation takes into account also systematics and it is discussed in Section \ref{sec:Data}.}\\
\hhline{=====}
Calendar date & MJD & Right Ascension & Declination & Angular separation\\
$[$UT$]$ & & &  & [arcsec]\\
\hline
\endfirsthead

\caption{Continued from previous page. Measured positions of RK1 and its angular separation from \maxithirt{}.}\\
\hhline{=====}
Calendar date & MJD & Right Ascension & Declination & Angular separation\\
$[$UT$]$ & & &  & [arcsec]\\
\hline
\endhead

\hline \multicolumn{5}{r}{{Continued on next page}}\\
\endfoot

\insertTableNotes
\endlastfoot
2019-03-09 & 58551.100 & 13$^{\rm h}$48$^{\rm m}$13.03$^{\rm s}$ $\pm$ 0.09\arcsec & $-63\degree$16\arcmin26.26\arcsec $\pm$ 0.20\arcsec & 2.80  $\pm$ 0.46\\    
2019-03-18 & 58560.075 & 13$^{\rm h}$48$^{\rm m}$13.11$^{\rm s}$ $\pm$ 0.02\arcsec & $-63\degree$16\arcmin25.31\arcsec $\pm$ 0.04\arcsec & 3.89  $\pm$ 0.44\\
2019-03-25 & 58567.074 & 13$^{\rm h}$48$^{\rm m}$13.19$^{\rm s}$ $\pm$ 0.04\arcsec & $-63\degree$16\arcmin24.45\arcsec $\pm$ 0.08\arcsec & 4.92  $\pm$ 0.44\\ 
2019-03-31 & 58573.759 & 13$^{\rm h}$48$^{\rm m}$13.29$^{\rm s}$ $\pm$ 0.19\arcsec & $-63\degree$16\arcmin23.57\arcsec $\pm$ 0.23\arcsec & 6.00  $\pm$ 0.20\\ 
2019-04-01 & 58574.061 & 13$^{\rm h}$48$^{\rm m}$13.23$^{\rm s}$ $\pm$ 0.42\arcsec & $-63\degree$16\arcmin23.83\arcsec $\pm$ 0.52\arcsec & 5.58  $\pm$ 0.49\\
2019-04-08 & 58581.715 & 13$^{\rm h}$48$^{\rm m}$13.363$^{\rm s}$ $\pm$ 0.005\arcsec & $-63\degree$16\arcmin22.77\arcsec $\pm$ 0.01\arcsec & 6.78  $\pm$ 0.02\\
2019-04-09 & 58582.053 & 13$^{\rm h}$48$^{\rm m}$13.31$^{\rm s}$ $\pm$ 0.41\arcsec & $-63\degree$16\arcmin23.51\arcsec $\pm$ 0.13\arcsec & 6.15  $\pm$ 0.48\\
2019-04-15 & 58588.053 & 13$^{\rm h}$48$^{\rm m}$13.36$^{\rm s}$ $\pm$ 0.04\arcsec & $-63\degree$16\arcmin22.43\arcsec $\pm$ 0.05\arcsec & 7.23  $\pm$ 0.44\\
2019-04-16 & 58589.806 & 13$^{\rm h}$48$^{\rm m}$13.40$^{\rm s}$ $\pm$ 0.02\arcsec & $-63\degree$16\arcmin22.13\arcsec $\pm$ 0.01\arcsec & 7.64  $\pm$ 0.25\\
2019-04-20 & 58593.074 & 13$^{\rm h}$48$^{\rm m}$13.44$^{\rm s}$ $\pm$ 0.08\arcsec & $-63\degree$16\arcmin21.97\arcsec $\pm$ 0.07\arcsec & 7.90  $\pm$ 0.44\\           
2019-04-29 & 58602.144 & 13$^{\rm h}$48$^{\rm m}$13.72$^{\rm s}$ $\pm$ 0.75\arcsec & $-63\degree$16\arcmin20.88\arcsec $\pm$ 0.15\arcsec & 9.39  $\pm$ 0.25\\
2019-04-30 & 58603.306 & 13$^{\rm h}$48$^{\rm m}$13.58$^{\rm s}$ $\pm$ 0.21\arcsec & $-63\degree$16\arcmin21.05\arcsec $\pm$ 0.04\arcsec & 9.14  $\pm$ 0.18\\
2019-05-04 & 58607.908 & 13$^{\rm h}$48$^{\rm m}$13.52$^{\rm s}$ $\pm$ 0.08\arcsec & $-63\degree$16\arcmin20.64\arcsec $\pm$ 0.28\arcsec & 9.20  $\pm$ 0.25\\
2019-05-11 & 58614.909 & 13$^{\rm h}$48$^{\rm m}$13.53$^{\rm s}$ $\pm$ 0.15\arcsec & $-63\degree$16\arcmin20.40\arcsec $\pm$ 0.37\arcsec & 9.69  $\pm$ 0.31\\
2019-05-18 & 58621.888 & 13$^{\rm h}$48$^{\rm m}$13.69$^{\rm s}$ $\pm$ 0.31\arcsec & $-63\degree$16\arcmin18.88\arcsec $\pm$ 0.81\arcsec & 11.61 $\pm$ 0.68\\
2019-10-19 & 58775.616 & 13$^{\rm h}$48$^{\rm m}$14.99$^{\rm s}$ $\pm$ 0.28\arcsec & $-63\degree$16\arcmin06.31\arcsec $\pm$ 0.15\arcsec & 26.74 $\pm$ 0.46\\         
2019-10-26 & 58782.597 & 13$^{\rm h}$48$^{\rm m}$14.94$^{\rm s}$ $\pm$ 0.25\arcsec & $-63\degree$16\arcmin06.28\arcsec $\pm$ 0.15\arcsec & 26.58 $\pm$ 0.46\\           
2019-11-01 & 58788.670 & 13$^{\rm h}$48$^{\rm m}$14.84$^{\rm s}$ $\pm$ 0.58\arcsec & $-63\degree$16\arcmin05.68\arcsec $\pm$ 0.14\arcsec & 26.73 $\pm$ 0.48\\           
2019-11-10 & 58797.408 & 13$^{\rm h}$48$^{\rm m}$14.99$^{\rm s}$ $\pm$ 0.28\arcsec & $-63\degree$16\arcmin05.47\arcsec $\pm$ 0.67\arcsec & 27.45 $\pm$ 0.71\\          
2019-11-18 & 58805.403 & 13$^{\rm h}$48$^{\rm m}$14.99$^{\rm s}$ $\pm$ 0.13\arcsec & $-63\degree$16\arcmin05.50\arcsec $\pm$ 0.36\arcsec & 27.41 $\pm$ 0.53\\          
2019-11-24 & 58811.345 & 13$^{\rm h}$48$^{\rm m}$14.97$^{\rm s}$ $\pm$ 0.18\arcsec & $-63\degree$16\arcmin06.92\arcsec $\pm$ 0.39\arcsec & 26.63 $\pm$ 0.41\\
2019-11-27 & 58814.722 & 13$^{\rm h}$48$^{\rm m}$14.82$^{\rm s}$ $\pm$ 0.78\arcsec & $-63\degree$16\arcmin06.05\arcsec $\pm$ 0.05\arcsec & 26.36 $\pm$ 0.26\\
2019-11-30 & 58817.449 & 13$^{\rm h}$48$^{\rm m}$14.90$^{\rm s}$ $\pm$ 0.25\arcsec & $-63\degree$16\arcmin06.27\arcsec $\pm$ 0.27\arcsec & 27.19 $\pm$ 0.31\\
2019-12-03 & 58820.734 & 13$^{\rm h}$48$^{\rm m}$15.16$^{\rm s}$ $\pm$ 0.79\arcsec & $-63\degree$16\arcmin06.74\arcsec $\pm$ 0.41\arcsec & 27.03 $\pm$ 0.43\\
2019-12-07 & 58824.396 & 13$^{\rm h}$48$^{\rm m}$14.85$^{\rm s}$ $\pm$ 0.50\arcsec & $-63\degree$16\arcmin07.20\arcsec $\pm$ 0.65\arcsec & 26.40 $\pm$ 1.36\\
2019-12-10 & 58827.754 & 13$^{\rm h}$48$^{\rm m}$14.93$^{\rm s}$ $\pm$ 0.17\arcsec & $-63\degree$16\arcmin05.48\arcsec $\pm$ 0.15\arcsec & 27.33 $\pm$ 0.24\\
2019-12-14\tnote{a} & 58831.316 & 13$^{\rm h}$48$^{\rm m}$15.05$^{\rm s}$ $\pm$ 0.16\arcsec & $-63\degree$16\arcmin05.88\arcsec $\pm$ 0.28\arcsec & 27.44 $\pm$ 0.33\\
2019-12-20 & 58837.428 & 13$^{\rm h}$48$^{\rm m}$14.93$^{\rm s}$ $\pm$ 1.19\arcsec & $-63\degree$16\arcmin05.28\arcsec $\pm$ 0.53\arcsec & 27.38 $\pm$ 0.69\\
2020-02-08 & 58887.185 & 13$^{\rm h}$48$^{\rm m}$15.03$^{\rm s}$ $\pm$ 0.23\arcsec & $-63\degree$16\arcmin05.23\arcsec $\pm$ 0.59\arcsec & 27.94 $\pm$ 0.56\\           
\hline
\label{tab:first_jet_angsep}
\end{longtable}
\end{center}
\end{ThreePartTable}

\section{Number of radiating electrons}
\label{sec:number of electrons}

We present here in detail the calculation of the number or radiating electrons $N_{\rm e}$ that we applied in Section \ref{sec:Jet composition}. This is obtained by performing standard minimum energy calculations, as outlined in \cite{Longair}. We consider a synchrotron source with volume $V$ and monochromatic luminosity $L_{\nu}$ at frequency $\nu$, characterised by a power law spectrum $L_{\nu} \propto \nu^{-\alpha}$ (we note that here we use a different notation compared to Section \ref{sec:Introduction} for the radio spectral index $\alpha$). We assume a standard electron distribution modelled as a power law with index $p = 2\alpha + 1$, where the electron energy spectrum per unit volume is

\begin{equation}
N(E) {\rm d} E = k E^{-p} {\rm d} E
\label{eq:spectrum}
\end{equation}

The synchrotron radio luminosity $L_{\nu}$ depends on the energy spectrum of relativistic electrons and on the magnetic field $B$ as
\begin{equation}
  L_{\nu} = A(\alpha) V k B^{1+\alpha} \nu^{-\alpha} 
\label{eq:radio_lum}
\end{equation}
for which the constant $A(\alpha)$ can be expressed as
\begin{equation}
A(\alpha) = \frac{\sqrt{3} e^3}{4\pi \epsilon_0 m_{\rm e} c}  \left[ \frac{3}{2}  \frac{e}{\pi m_{\rm e}^3 c^4}         \right]^{\frac{p-1}{2}} a(p)
\end{equation}
where $\epsilon_0$ is the permittivity of free space, and for which
\begin{equation}
a(p) = \frac{\sqrt{\pi}}{2}  \frac{ \Gamma\left( \frac{p}{4} + \frac{19}{12}  \right) \Gamma\left( \frac{p}{4} - \frac{1}{12}  \right) \Gamma\left( \frac{p}{4} + \frac{5}{4}  \right)   }{(p+1) \Gamma\left( \frac{p}{4} + \frac{7}{4}  \right)}
\end{equation}
where $\Gamma(x)$ is the Gamma function.
 
To obtain the total number of relativistic electrons, we integrate Equation \ref{eq:spectrum} between the minimum and maximum electron energies
\begin{equation}
N_e = V \int^{E_{\rm max}}_{E_{\rm min}} N(E) dE = V\int_{E_{\rm min}}^{E_{\rm max}} k E^{-p} dE = \frac{\nu^{\left(\frac{p-1}{2}\right)}}{p-1} \left( \frac{1}{C} \right)^{\frac{1-p}{2}} \left[ \frac{L_{\nu}}{A(\alpha) B}    \right] \left[ \nu_{\rm min}^{\frac{1-p}{2}} - \nu_{\rm max}^{\frac{1-p}{2}} \right] 
\end{equation}
for which we express $E_{\rm min}$ and $E_{\rm max}$ as a function of the minimum and maximum frequencies at which the electrons radiate
\begin{equation}
E_{\rm min} = \left( \frac{\nu_{\rm min}}{CB}   \right)^{\frac{1}{2}}, \ \ \ E_{\rm max} = \left( \frac{\nu_{\rm max}}{CB}   \right)^{\frac{1}{2}}
\end{equation}
where
\begin{equation}
C = 0.29 \cdot \frac{3}{2} \frac{e}{2\pi m_{\rm e}} \frac{1}{(m_{\rm e} c^2)^2}
\end{equation}

At equipartition conditions, the magnetic field $B $ corresponding to the minimisation of the total energy can be written as
\begin{equation}
B = \left[ \frac{3\mu_0}{2}   \frac{G(\alpha)\eta L_{\nu}}{V}   \right]^{\frac{2}{7}}
\label{eq:magnetic_field_appendix}
\end{equation}
where $\mu_0$ and $\eta$ are discussed in Section \ref{sec:Jet composition}, and where
\begin{equation}
G(\alpha) = \frac{1}{a(p)} \frac{1}{(p-2)} \left[ \nu_{\rm min}^{-(\frac{p-2}{2})}     - \nu_{\rm max}^{-(\frac{p-2}{2})}    \right]  \nu^{\frac{p-1}{2}} C^{-(\frac{p-2}{2})}  \frac{4\pi \epsilon_0 m_{\rm e} c}{\sqrt{3} e^3} \left[ \frac{3}{2}  \frac{e}{\pi m_{\rm e}^3 c^4}         \right]^{\frac{1-p}{2}}
\end{equation}

\section{Additional figures}
\label{Alternative models}

\begin{figure*}
\begin{center}
\includegraphics[width=0.8\textwidth]{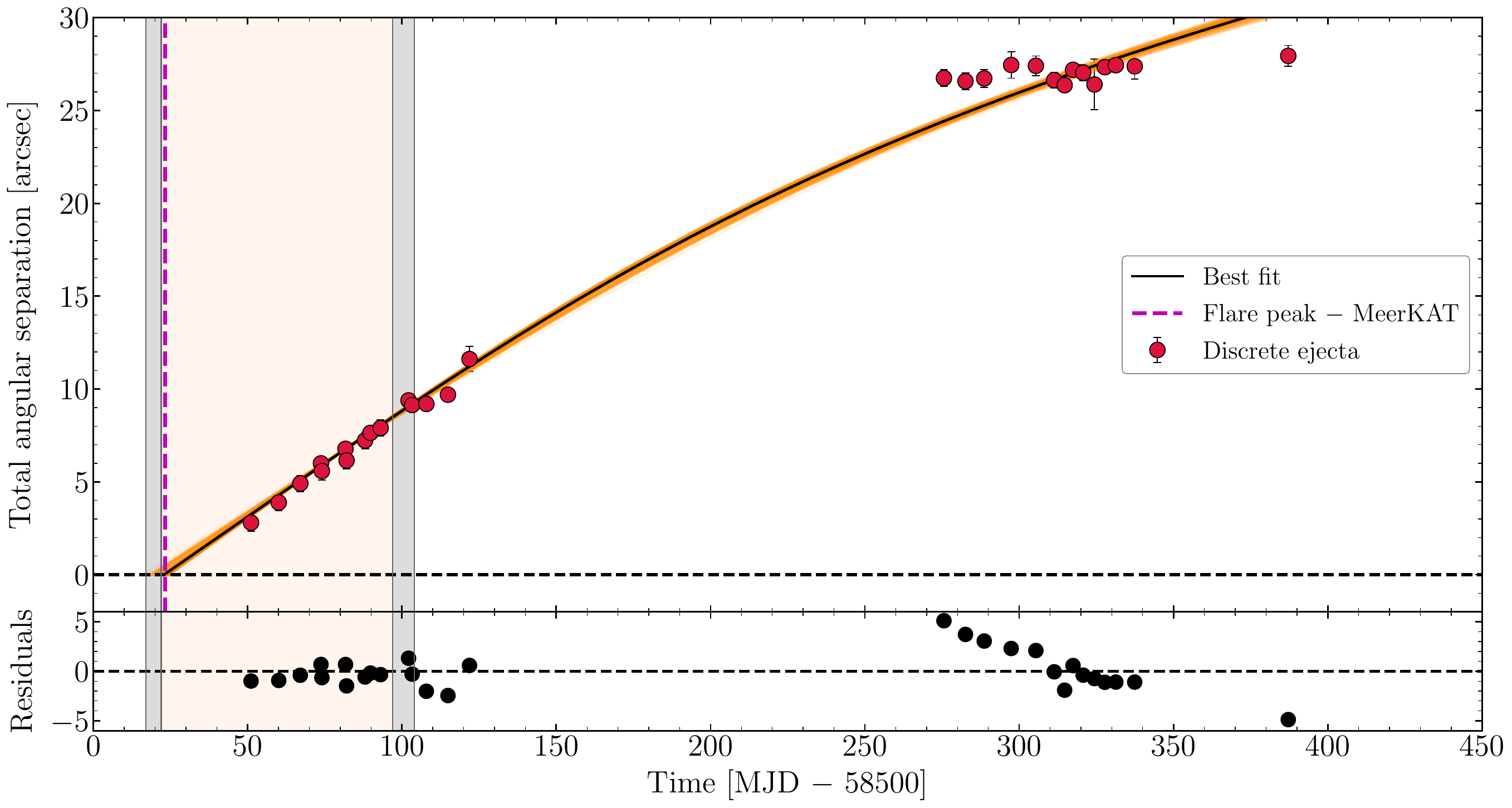}
\caption{Same as Figure \ref{fig:angsep_fit_uncertainties}, but assuming an ISM with uniform density. As the model without the inclusion of a cavity fails to reproduce accurately the jet deceleration phase, we deem this scenario unlikely.}
\label{fig:angsep_fit_uncertainties_nocavity}
\end{center}
\end{figure*}

We show in this section the results for the two other possible parameter choices discussed
in Section \ref{sec:Results}. The first possibility involves a jet moving in a uniform ISM with density $n_{\rm ism} = 1$ cm$^{-3}$ (fixing $\delta = 1$ and $R_{\rm c} = 0$). The fit is shown in Figure \ref{fig:angsep_fit_uncertainties_nocavity}. As mentioned in Section \ref{sec:Uniform density}, this scenario does not appear consistent with our data, as the model fails to reproduce the decelerating part of the jet motion.

\begin{figure*}
\begin{center}
\includegraphics[width=0.8\textwidth]{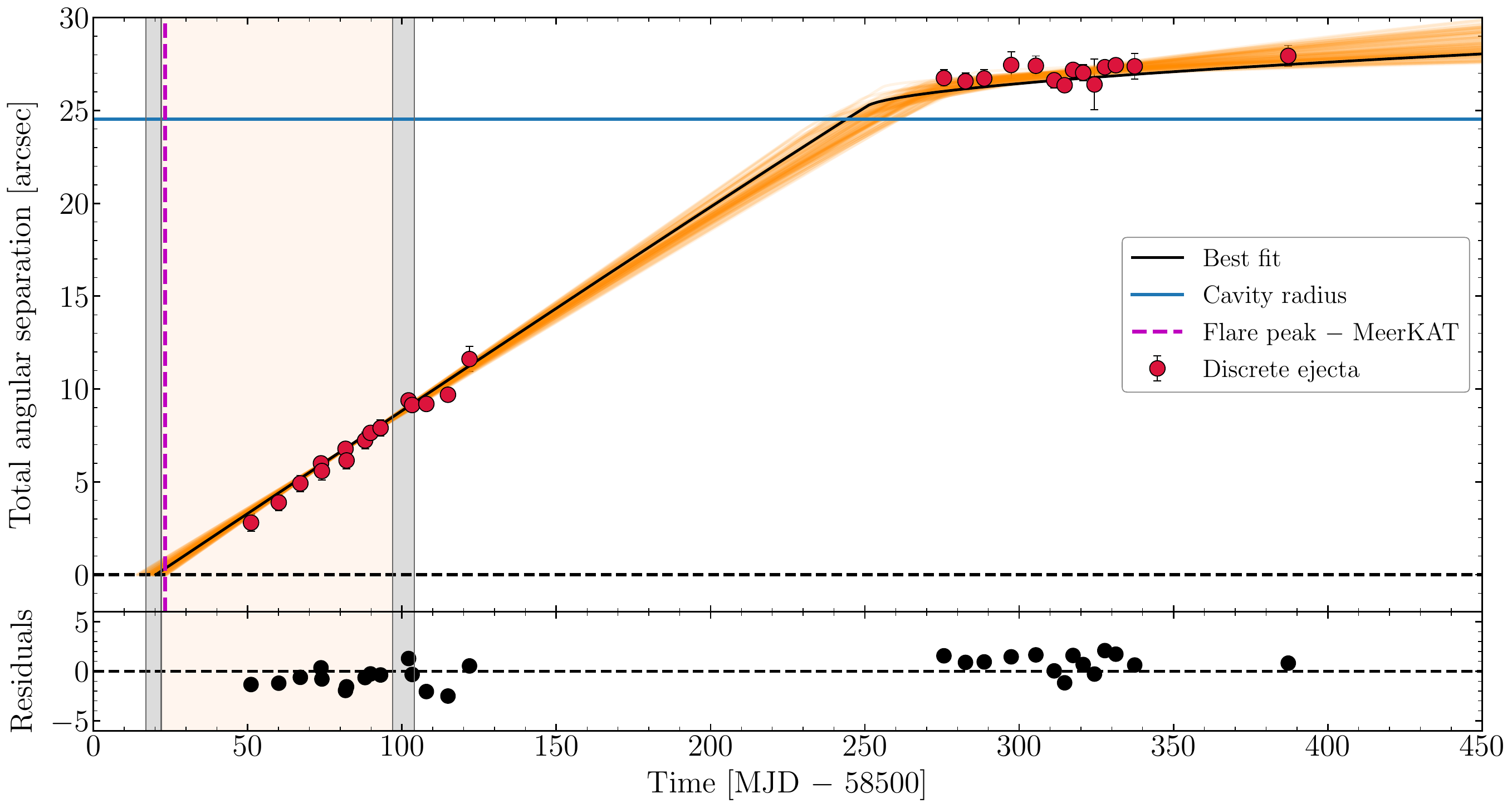}
\caption{Same as Figure \ref{fig:angsep_fit_uncertainties}, but assuming a prior on the source distance centered on 3.39 kpc \protect\citep{Lamer_2021}.
The model appears to fit reasonably well our data, and the results are very similar to what reported in Section \ref{sec:Motion in a low-density ISM cavity}, implying that the the choice of the prior on the source distance does not significantly affect the conclusions of this work.}
\label{fig:angsep_fit_uncertainties_erosita}
\end{center}
\end{figure*}

The second possibility, presented in Section \ref{sec:Dust-scattering halo distance}, consists of fitting the jet motion including the presence of a low-density ISM cavity, with the same setup as the main results presented in this work, in Section \ref{sec:Motion in a low-density ISM cavity} and in the Discussion. The only difference is in the prior on the source distance, that now is chosen to be centered at 3.39 kpc, a value obtained by \cite{Lamer_2021} from the detection with eROSITA of X-ray echoes from a giant dust-scattering ring. The fit is shown in Figure \ref{fig:angsep_fit_uncertainties_erosita}. The plot appears very similar to the main one, which shown in Figure \ref{fig:angsep_fit_uncertainties}, implying that the the choice of the prior on the source distance does not significantly affect the conclusions of this paper.

% Don't change these lines
\bsp	% typesetting comment
\label{lastpage}
\end{document}